\pdfoutput=1
\documentclass[twocolumn,english,aps,superscriptaddress,prb]{revtex4-1}

\usepackage{babel}
\usepackage{amsmath}
\usepackage{amssymb}
\usepackage{graphicx}
\usepackage{color}

\begin{document}

\title{Excitation spectrum and Density Matrix Renormalization Group iterations}

\author{Natalia Chepiga}
\affiliation{Institute of Physics, Ecole Polytechnique F\'ed\'erale de Lausanne (EPFL), CH-1015 Lausanne, Switzerland}
\author{Fr\'ed\'eric Mila}
\affiliation{Institute of Physics, Ecole Polytechnique F\'ed\'erale de Lausanne (EPFL), CH-1015 Lausanne, Switzerland}

\date{\today}
\begin{abstract} 
We show that, in certain circumstances, exact excitation energies appear as locally site-independent (or flat) modes if one records the excitation spectrum 
of the effective Hamiltonian while sweeping through the lattice in the variational Matrix Product State formulation of the Density Matrix 
Renormalization Group (DMRG), a remarkable property since the effective Hamiltonian is only constructed to target the ground state. 
Conversely, modes that are very flat over several consecutive iterations
are systematically found to correspond to faithful excitations. We suggest to use this property to extract accurate information about excited 
states using the standard ground state algorithm. The results are spectacular for critical systems, for which the 
low-energy conformal tower of states can be obtained very accurately at essentially no additional cost, 
as demonstrated by confirming the predictions of boundary conformal field theory for two simple minimal models - the transverse-field Ising 
model and the critical three-state Potts model. This approach is also very efficient to detect the quasi-degenerate low-energy excitations in topological
phases, and to identify localized excitations in systems with impurities. Finally, using the variance of the Hamiltonian as a criterion, we assess the 
accuracy of the resulting Matrix Product State representations of the excited states.
\end{abstract}

\maketitle


 \section{Introduction}
 
Modern quantum many-body physics relies to a large extent on numerical simulations. Over the years, the Density Matrix Renormalization Group (DMRG)\cite{dmrg1,dmrg2} has established itself as the most powerful tool for strongly correlated one-dimensional systems. The reformulation of the algorithm in terms of Matrix Product States (MPS) \cite{dmrg3,dmrg4} has not only brought new insights in one dimension, but has also boosted the generalization of the DMRG to higher dimensions, an approach known as Tensor Network algorithms.
 
The DMRG algorithm has first been developed as a method to search for the ground-state. To access excited states is in general significantly more involved. Several approaches have been developed to address this problem.

If some symmetry can be imposed on the wave-function, and if the excited state of interest is the lowest-energy state of some symmetry sector, the search for this excited state is then simply a ground-state search within the corresponding symmetry sector. In particular, this approach leads to a rather straightforward calculation of the singlet-triplet gap in quantum antiferromagnets by calculating the lowest energy states in the sectors of total magnetization $S^z_\mathrm{tot}=0$ and $S^z_\mathrm{tot}=1$. 

If however symmetry cannot be used to distinguish excitations, for instance if the Hamiltonian does not preserve any symmetry, or if the excitation lies in the symmetry sector of the ground state, the algorithm has to be modified significantly.
In conventional DMRG, the way to proceed is to construct the density matrix not only from basis vectors that appear in the Schmidt decomposition of the ground state, but also from basis vectors that appear in the Schmidt decomposition of low-lying excitations\cite{dmrg5,chandross,bursill,ortolani,mcculloch}. Typically one targets five or fewer excited states \cite{dmrg2}. This approach can also be implemented in variational MPS,  using tensors that encode mixed states. All excited states are then calculated together with the ground-state, but the price to pay is high since the bond
dimension increases very fast with the number of eigenstates.

The MPS representation allows for a significant improvement in that respect: After the construction of the ground-state, one can search for an eigenstate that is orthogonal to the ground state and has the smallest energy\cite{dmrg4,PhysRevB.73.014410,mcculloch}. Higher excitations can also be accessed by looking for an eigenstate that is orthogonal to all previously constructed eigenvectors of the Hamiltonian. This method is systematic and well controlled, but the algorithm has to be re-run for each eigenstate. Moreover, the states should be well converged, otherwise the error will accumulate in the following runs. So it becomes very heavy if one wishes to access
many eigenstates. 

In the present paper, we show that, in certain cases, there is a much cheaper alternative. It relies on keeping track of several eigenvalues of the 
effective Hamiltonian during DMRG iterations, usually referred to as sweeps. In general, the Hamiltonian written in the effective basis of the ground state gives only a poor estimate of the excitation spectrum. However, in some particular cases, this effective Hamiltonian gives access to very accurate estimates of 
excitation energies, as we shall demonstrate in three contexts:  {\it i)} critical systems; {\it ii)} in-gap states in topological phases; {\it iii)} localized excitations. 
The central observation is that, in all these cases, exact excitation energies are essentially completely flat during part of the sweeps. Quite remarkably, the flat behavior does not necessarily occur at the center of the system, where the algorithm is expected to be most accurate, but it can also occur near the edges or close to the impurity.

The calculation of the excitation spectra of critical systems is perhaps the most important application of this method. According to conformal field theory, the excitation spectrum at a critical point forms a conformal tower that is characteristic of the universality class of the transition and of the underlying critical theory. 
Traditionally, the universality class of the transition is identified by computing the central charge and the critical exponents of some observables (on-site magnetization, spin-spin correlations, etc.). The determination of the critical exponents is however extremely sensitive to logarithmic corrections if they are 
present, and the calculation of the central charge can be affected by strong finite-size effects. Quite generally, the excitation spectrum contains more information on the underlying critical theory. It can be used as a useful complement to the central charge and the critical exponents, and in cases where those are difficult to extract, the conformal tower becomes the method of choice \cite{J1J2J3_letter,J1J2Jb_comment}. 
 

The paper is organized as follows. In Section \ref{sec:intro_MPS}, we briefly review the basics of the Matrix Product State formulation of DMRG, and we put forward the central idea of this paper about the possibility of extracting excitation energies from DMRG iterations. In Section \ref{sec:excitspec}, we benchmark the method on the transverse field Ising model, for which the full spectrum of open chains with free boundary conditions can be calculated exactly using Jordan-Wigner transformation.
In Section \ref{sec:minmod_ising}, we apply this method to the critical transverse field Ising model. By changing the boundary conditions of open chains, we show that the special structure of the  conformal towers of all the primary fields that appear in the Ising minimal model can be obtained along these lines, in perfect agreement with the predictions of boundary conformal field theory (CFT). Section \ref{sec:minmod_potts} is dedicated to a similar study of another minimal model - the three-state Potts model.  In Section \ref{sec:edge}, we show that the gap between the low-lying in-gap states can be extracted in the same way, and we show how the algorithm can be modified to detect the presence of the edge states at very low computational cost.
In Section \ref{sec:local}, we show that it is also possible to detect localized excitations around impurities. The accuracy of the resulting MPS representation of the excited states is discussed in Section \ref{sec:accuracy}.
The results are summarized and put in perspective in Section \ref{sec:conclusion}.

\section{The method}
\label{sec:intro}

\subsection{Introduction to MPS}
\label{sec:intro_MPS}

We start with a brief reminder on MPS notations and on the variational optimization of the ground-state (DMRG). 
A complete and pedagogical introduction to the algorithm can be found in a recent review\cite{dmrg4}. 

Let us consider a chain of $N$ interacting spins-$S$. The Hilbert space of the chain grows exponentially fast with the number of sites as $d^N$, where $d$ is the size of the local Hilbert space $d=2S+1$. The limitation on the memory restricts the maximal number of sites for which the quantum state can be written explicitly to $N\approx 20$. 
  The key point of the MPS representation is to overcome this restriction and to write the state as a product of local tensors. This can be done since any quantum state of a bipartite system can be effectively represented in a compact basis constructed with the Schmidt decomposition. According to linear algebra, for any rectangular matrix $M$ of dimension $m\times n$ there exists a singular value decomposition (SVD) $M=USV^\dagger$, where $U$ is of dimension $m\times\min(m,n)$ and is left-normalized ($U^\dagger U=I$); $V$ is of dimension $n\times\min(m,n)$ and is right-normalized ($VV^\dagger=I$); $S$ is a diagonal matrix of dimension $\min(m,n)$ with non-negative entries.
   
\begin{figure}[h]
\centering
\includegraphics[width=0.49\textwidth]{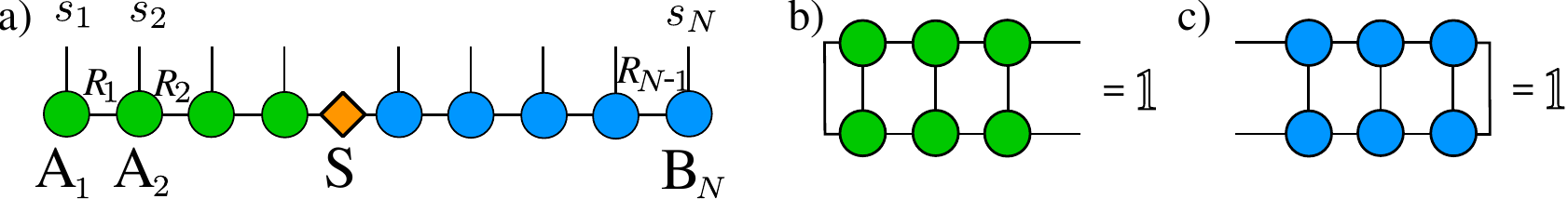}
\caption{(Color online) Graphical notations for Matrix Product State representation. (a) MPS representation of a state $|\psi\rangle$ in terms of left- (green) and right-normalized (blue) tensors and of the diagonal matrix $S$ whose entries are given by the Schmidt values. The tensors are contracted over connected bonds. The vertical unconnected bonds correspond to the physical (spin) index. (b) and (c) Graphical representation of left and right normalizations.  }
\label{fig:mps1}
\end{figure}

Using successive SVD decompositions, any quantum state can be represented in terms of local three-dimensional tensors. One leg of each tensor corresponds to the physical (spin) index of dimension $d$. 
The remaining legs are auxiliary: a tensor is connected to its left and right neighbors by contracting the corresponding auxiliary bonds. In practice, the state is given in the mixed-canonical representation including both left- ($A_i$) and right-normalized ($B_i$) tensors, as shown in Fig.\ref{fig:mps1}(a). The graphical representation of the normalization condition is shown in Fig.\ref{fig:mps1}(b-c). Connected lines correspond to the contracted bonds of the tensors.

\begin{figure}[t]
\centering
\includegraphics[width=0.49\textwidth]{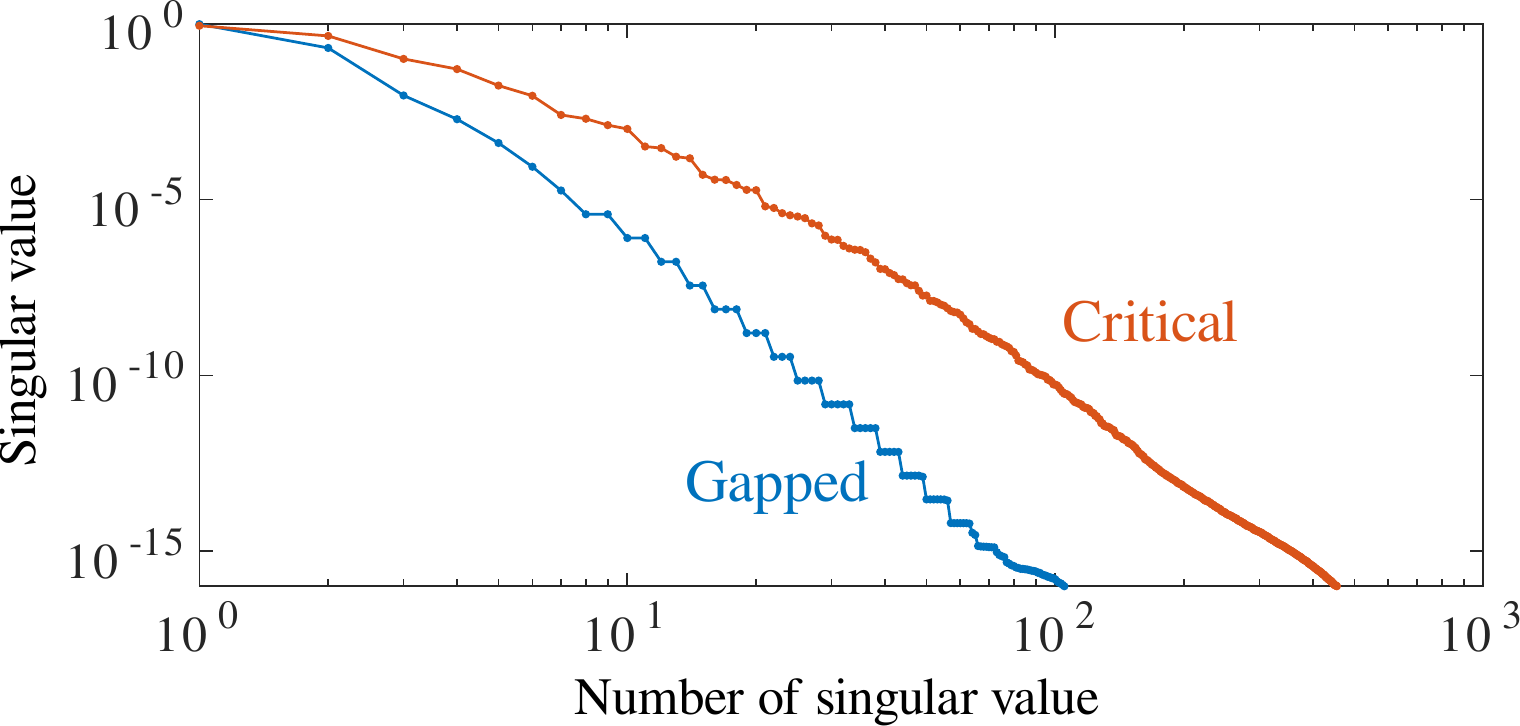}
\caption{(Color online)  Decay of the Schmidt values for gapped and critical systems. Here we show singular values computed in the middle of the spin-1/2 transverse field Ising chain with $N=100$ spins at the critical point $h=0.5J$ and inside the gapped phase $h=0.7J$.}
\label{fig:schmidt}
\end{figure}

The auxiliary bond dimension $R_j$ grows exponentially with the distance to the edges as $R_j=\min(d^j,d^{N-j})$. So the decomposition itself does not reduce the required amount of memory. However, for strongly correlated systems, the Schmidt values $S_{i,i}$ decay fast with $i$, as shown in Fig.\ref{fig:schmidt}. Therefore the exact decomposition of the matrix can be replaced by an approximate one:
$M_{k,l}\approx\sum_{i=1}^{D} U_{k,i}S_{i,i}V^\dagger_{i,l}$,
where the summation index $i$ only  runs over a few of the largest Schmidt values $D<R_j=\min(m,n)$. With this approximation, the bond dimension $R_j$ of the MPS representation is given by $D_j=\min(d^i,d^{N-i},D)$. The number $D$ is known in the literature as 'the number of kept states'. According to the area law,  this number is a constant  that does not depend on the system size for gapped systems. Typically it varies between a few hundreds and a few thousands, which is much smaller than the size of the total Hilbert space. In critical systems, the decay of Schmidt values is slower (see Fig.\ref{fig:schmidt} for comparison), and $D$ depends on the system size.

The search for the ground state consists in finding an MPS representation of the state $|\psi\rangle$ that minimizes the energy:
\begin{equation}
  E=\frac{\langle\psi | H|\psi\rangle}{\langle\psi |\psi\rangle}
\end{equation}
The Hamiltonian is itself represented as a Matrix Product Operator (MPO), a product of local matrices with two auxiliary indices and two physical indices\cite{dmrg4}. The algorithm operates iteratively: While all but two (or a few) tensors of the MPS are kept constant, a selected couple of tensors are updated by solving the corresponding eigenvalue problem.

\begin{figure}[t]
\centering
\includegraphics[width=0.49\textwidth]{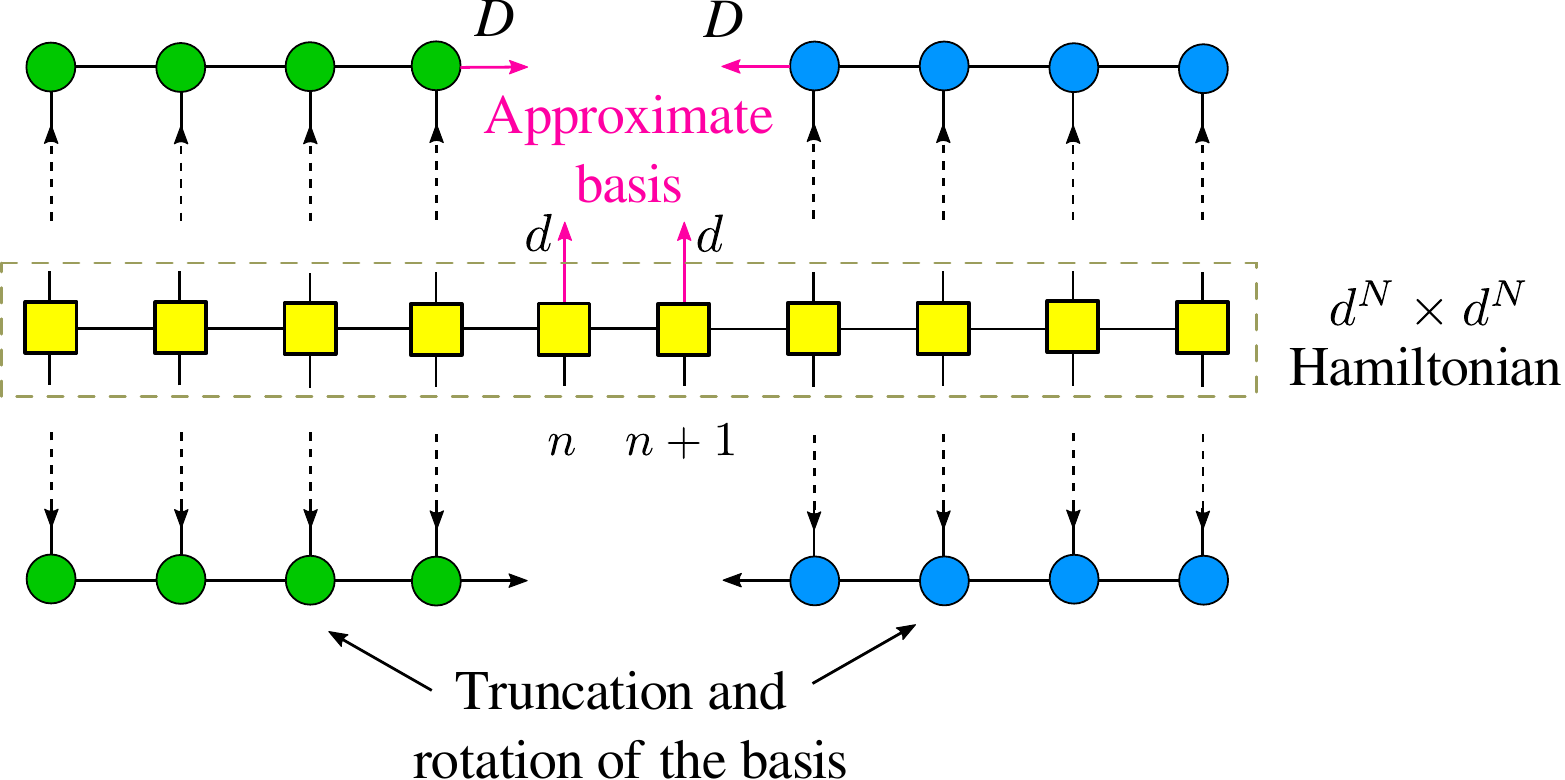}
\caption{(Color online)  Graphical representation of the total and effective Hamiltonians: The row of on-site MPO (yellow boxed) contracted through the auxiliary bonds is a $d^N\times d^N$ matrix that represents the {\bf total} Hamiltonian. Contracting through all bonds except the physical indices at sites $n$ and $n+1$ gives rise to a local effective Hamiltonian
that acts in a Hilbert space of dimension $(dD)^2\ll d^{N}$. The collection of $A_n$ and $B_n$ tensors approximate the basis via rotation and truncation. }
\label{fig:mps4}
\end{figure}
\begin{figure}[t]
\centering
\includegraphics[width=0.49\textwidth]{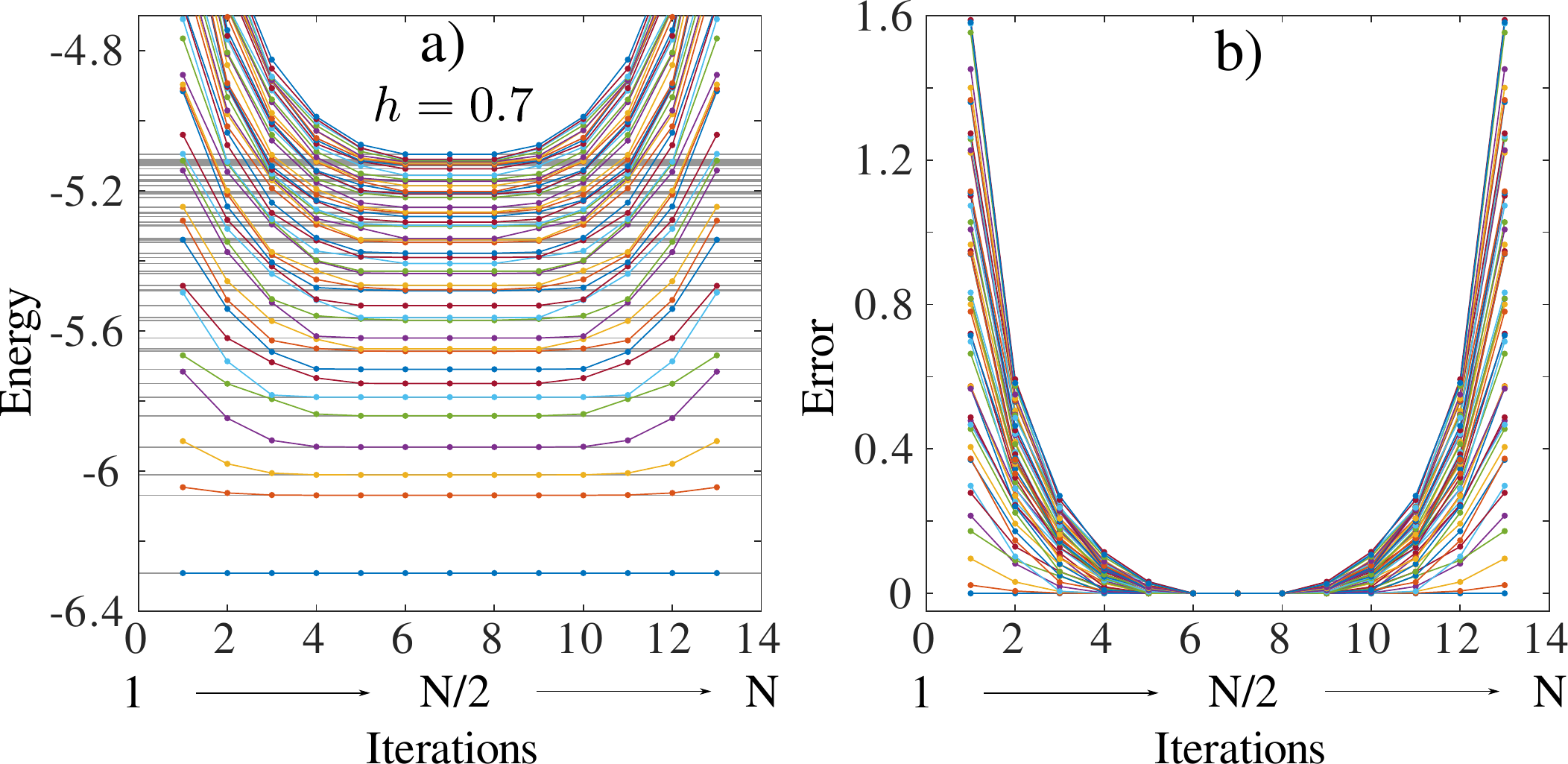}
\caption{(Color online)  Results of the diagonalization of the effective Hamiltonian with all $D=256$ kept states for the transverse field Ising chain of Eq.\ref{eq:isingmod} with $J=-1$ and $h=0.7$, in the gapped phase with a non-degenerate ground state. (a) Energy of the lowest 50 states as a function of iteration during left-to-right half-sweep. The exact results are provided for reference and shown with grey lines. (b) The difference between the energy obtained by diagonalizing the effective Hamiltonian and the exact values of the energy as a function of iteration. The difference vanishes for three points in the middle of the chain for all energy levels.}
\label{fig:ising_exact}
\end{figure}

Let us first consider the full MPS with all states kept so that the auxiliary bond of the MPS has dimension $R_j=\min(d^j,d^{N-j})$. The size of the effective Hamiltonian obtained by contraction of the MPO and MPS around sites $n$ and $n+1$ (as shown in Fig.\ref{fig:mps4}) is equal to $\min(d^{n-1},d^{N-n+1})\times \min(d^{n+1},d^{N-n-1})\times d^2$. This means that at the center of the chain for $n=[N/2-1,N/2,N/2+1]$ if $N$ is even and for $n=[(N-1)/2,(N+1)/2]$ if $N$ is odd the size of the Hilbert space in which the effective Hamiltonian operates is equal to the size of the total Hilbert space of the system. Thus the MPS contracted with the MPO does not truncate the set of basis vectors but only rotate the basis in which the Hamiltonian is written. This implies that, by keeping all states in the MPS, {\it i)} one can compute the complete spectrum by diagonalizing the effective Hamiltonian for sites $n=N/2-1$, $N/2$ or $N/2+1$ when $N$ is even or for sites $n=(N-1)/2$ and $n=(N+1)/2$ when $N$ is odd, and {\it ii)} since this calculation is exact, the values of the energies at these points are equal. Thus the energies as a function of iterations are completely flat in the middle of the chain, as can be seen in Fig.\ref{fig:ising_exact}. Note that the above statements hold true for both gapless and gapped systems. Due to the limitation of numerical precision, highly local Hamiltonians, e.g. chains of weakly coupled dimers, are exceptions. For instance, for spin-1/2 chain with alternating bonds, the coupling between the dimers should exceed $J_\mathrm{weak}/J_\mathrm{strong}\geq 10^{-3}$ in order to compute the exact spectrum for $N=16$ spins.  This comes from the fact that, for highly local Hamiltonians, the Schmidt values decrease immediately to almost-zero values that are of the order of machine precision, and thus the vectors that correspond to these states have nearly zero weight in the basis of the effective Hamiltonian. Interestingly, a few (4-5) excited states can still be captured within machine precision even for $J_\mathrm{weak}/J_\mathrm{strong}=10^{-6}$.

By truncating the bond dimension in SVD, we reduce the number of basis vectors. The basis formed by on-site states is no longer optimal. Instead, linear combinations of the basis vectors should be used. The tensors $A$ and $B$ constructed at each step perform a basis rotation in such a way that the new basis becomes the best set of basis vectors for the selected state - the ground state. Thus, the effective Hamiltonian diagonalized at each iteration can be understood as the original Hamiltonian written in a rotated and truncated basis of dimension $(Dd)^2$ (see Fig.\ref{fig:mps4}).

In terms of the truncated basis vectors, the slow decay of the singular values in critical systems means that more vectors possess essentially non-zero weight in the basis. Then more physical states can be efficiently constructed from the chosen set of basis vectors. Thus, one might expect that the truncated basis can be good enough to describe a few low-lying excited states, and the diagonalization of the original Hamiltonian written in this truncated basis could provide good estimates of the corresponding excitation energies. Moreover the size of the effective basis $(Dd)^2$ remains the same along the chain except very close to the edges. Then intuitively, if the size of the basis is sufficient to describe the excited state, the energy of this state will be flat while iterating along the chain, in complete analogy with the flattening of the exact spectrum in the three middle points when all states are kept. The rest of the paper is devoted to testing this
simple idea in a number of systems ranging from critical systems to gapped topological phases.

Throughout the paper, and unless specified otherwise, the simulations will be done with a variable bond dimension $D$ that increases during the simulation. We start with relatively small bond dimension in the range $22<D_\text{inf}<44$ in infinite-size DMRG, and we increase it by a factor 2.25 during the warm up. Therefore, after a warm-up, the bond dimension is in the range $50<D_\text{start}<100$. Then we increase the bond dimension linearly with each half-sweep up to its maximal value $D_\text{max}$, that will be specified for each model.  Regarding the Lanczos diagonalization\cite{lanczos} of the effective Hamiltonain, we typically perform at most $D+10n$ Lanczos iterations when we diagonalize the effective Hamiltonian if we target $n$ low-lying states.


\subsection{Excitation spectrum of the effective Hamiltonian: a simple test case}
\label{sec:excitspec}

In order to test this idea, we consider as a toy model the transverse field Ising model:

\begin{equation}  
H=-J\sum_i S^x_i S^x_{i+1}+hS^z_i,
\label{eq:isingmod}
\end{equation}
where $S^{x,z}_i$ are spin-1/2 operators at site $i$. A positive (resp. negative) coupling constant $J$ corresponds to the antiferromagnetic (resp. ferromagnetic) Ising model. In both cases, a quantum phase transition occurs at the critical values of the magnetic field $h=\pm J/2$.  The underlying conformal field theory will be discussed in details in the next section.

\begin{figure}[t]
\centering
\includegraphics[width=0.49\textwidth]{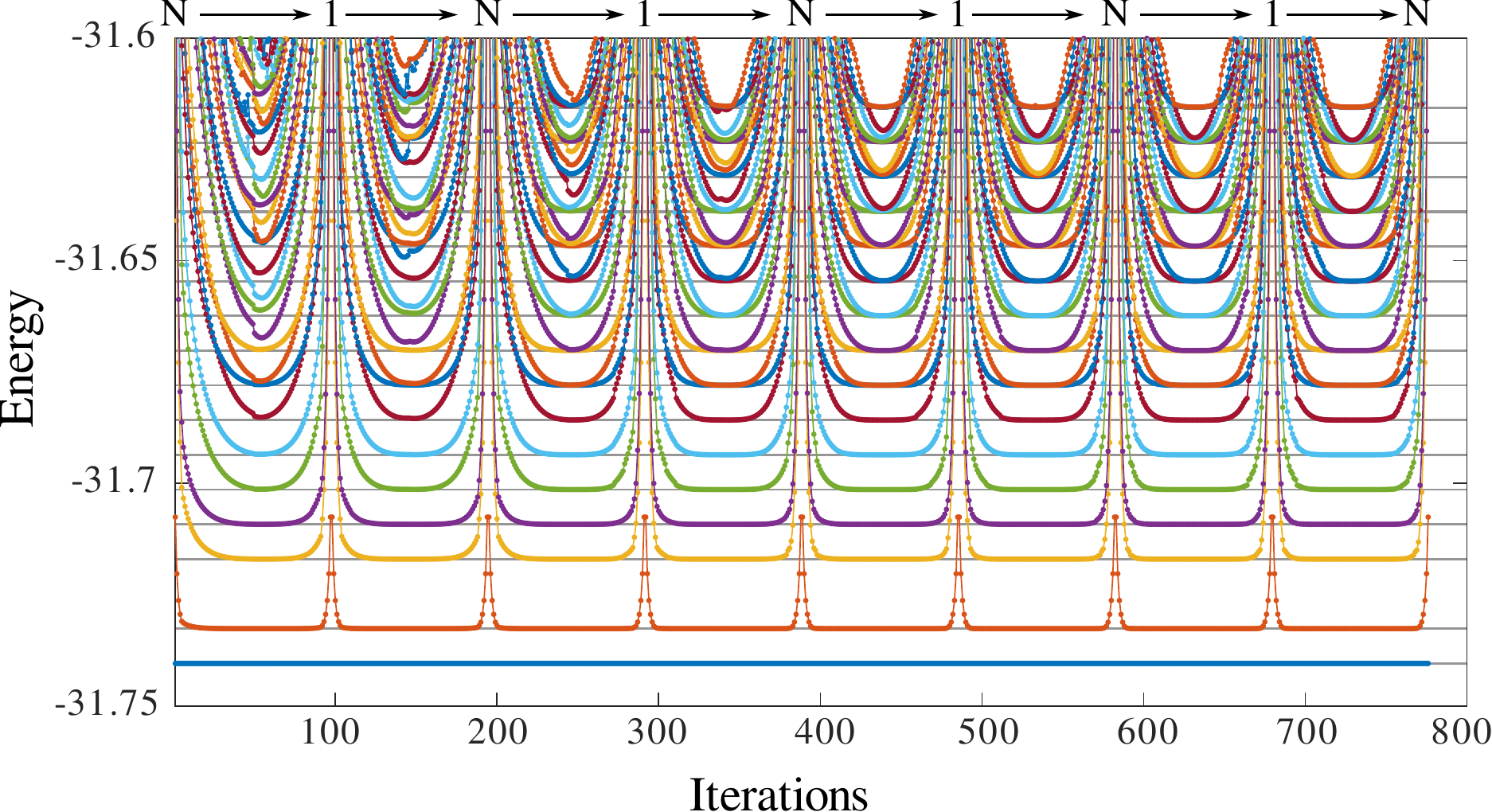}
\caption{(Color online) Energy of the 30 low-energy states in the critical Ising model in a transverse field with $N=100$ sites as a function of iterations. The flattening of the energies in the middle of the chain is an indicator of convergence. The periodic increase of the energy that occurs close to the chain boundary is the result of the reduced Hilbert space by MPS construction. Exact results are provided for reference and shown with grey lines. The plot starts with the first full sweep. The result from infinite-size DMRG and warm-up are not shown. }
\label{fig:ising_conv}
\end{figure}

\begin{figure}[h!]
\centering
\includegraphics[width=0.49\textwidth]{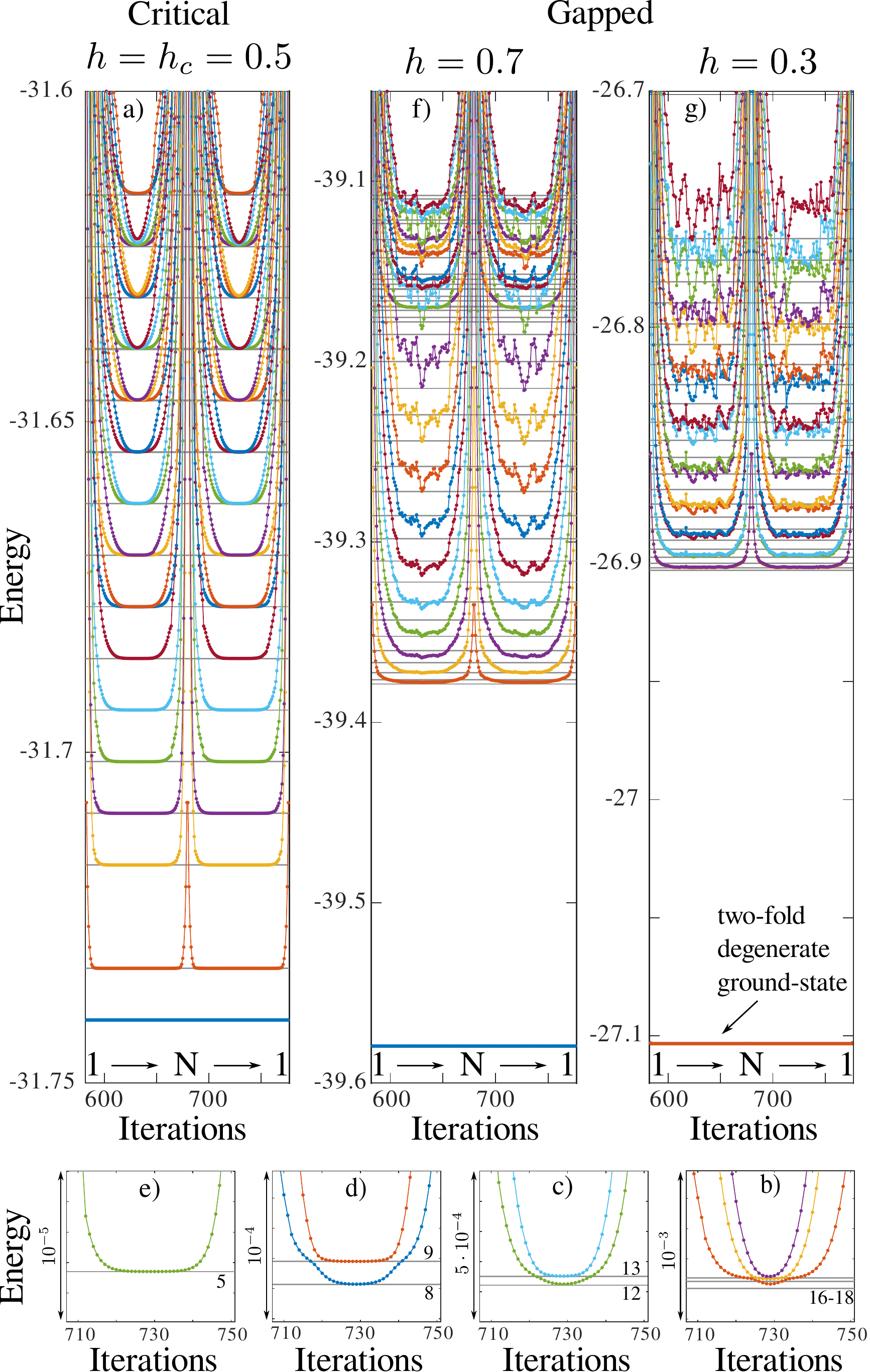}
\caption{(Color online) Excitation energies obtained during one DMRG sweep in the critical (a)-(e) and gapped (f),(g) transverse field Ising model for an open chain with $N=100$ sites. Exact results are provided for reference and shown with grey lines. (a) DMRG results at the center of the chain agree with exact energies. (b)-(e) Enlarged parts of (a) around different excitation levels. The ordinal numbers of the states are indicated in each plot, 1 corresponding to the ground-state. The number at the left side of plots (b)-(e) indicates the energy window around the selected energy level. (f) and (g) Far from criticality, DMRG results do not reproduce the exact energy levels: some states are missing, and strong oscillations appear.}
\label{fig:ising_compare}
\end{figure}

On a finite chain with open boundary conditions, the Hamiltonian (\ref{eq:isingmod}) can be rewritten as a quadratic form of Fermi operators using Jordan-Wigner transformation. Following Ref.\onlinecite{lieb,pfeuty}, the eigenvalue problem in the Hilbert space of dimension $2^N$ can be reduced to the diagonalization of an $N\times N$ matrix to find the elementary excitations. This can be performed exactly even for very long chains. The full spectrum can then be obtained by combining elementary excitations. The resulting spectrum is taken as a reference for the comparison with the DMRG results (gray lines in Fig.\ref{fig:ising_conv} and Fig.\ref{fig:ising_compare}).

When performing DMRG simulations for the transverse field Ising model, we typically set $D_\text{inf}=22$, $D_\text{start}=50$ and $D_\text{max}=200$. According to Fig.\ref{fig:mps4}, this corresponds to a truncation error of about $10^{-14}$ or lower. We performed up to $500$ Lanczos iterations while diagonalizing the Hamiltonian. An insufficient number of Lanczos iterations results in  some noise that affects the upper levels of the energy spectrum, as discussed in Appendix \ref{sec:noisy_ising}.

In Fig.\ref{fig:ising_conv}, we show the 30 lowest energy states of the critical Ising model computed by Lanczos diagonalization \cite{lanczos} of the effective Hamiltonian in the DMRG calculation of the ground-state as a function of the sweeps. This spectrum exhibits several typical remarkable features. First of all, in the middle of the chain, several levels are completely flat during a half sweep. 
By contrast, a sharp increase of all energies except the ground state energy occurs close to the chain boundary. This comes from the fact that the Hilbert space is too small at the edges by MPS construction. Secondly, the interval over which a given level is flat increases in subsequent sweeps, while the energy itself does not change significantly. Thirdly, more and more levels become flat upon increasing the number of sweeps. Finally, and most importantly, the energies that are stabilized in the flat portions match very accurately the exact results obtained with the Jordan-Wigner transformation, with no state missing, so that the low-energy part of the conformal tower is very well reproduced by our DMRG calculation. Even when states are quasi-degenerate, they are properly captured (see Fig.\ref{fig:ising_compare} b-e)).

By contrast, this is not the case when the system is not critical. In Fig.\ref{fig:ising_compare}, we compare the results obtained at the critical point with those obtained away form the critical point, below $h_c$, where the spectrum is gapped with a two-fold degenerate ground state, and above $h_c$, where the spectrum is gapped with a non-degenerate ground state. The results obtained for these gapped spectra differ from those obtained for the critical system in 
many respects. First of all, apart from the in-gap state below $h_c$, the energy levels are never really flat, apart maybe from the first one, but even that one
has a residual curvature definitely larger than those of the flat states of the critical case. Secondly, most of the levels are very noisy, and this noise is intrinsic, i.e. not due to the Lanczos truncation. Thirdly, the comparison with the exact spectrum
is rather poor. Many states are missing, and the energies around the center are in general not really close to an exact eigenstate.

We think that the difference comes from the very different nature of the excitations in both cases. In gapped phases, low-lying states can be significantly different from each other and can even belong to different symmetry (here parity) sectors, requiring essentially different sets of basis vectors. By contrast, at the critical point, the excitation spectrum is obtained by applying some primary fields and descendants on the ground state. Therefore it is natural that the basis obtained from the ground state also describes the excited states with high accuracy. This hypothesis agrees with the observation made by L{\"a}uchli in Ref.\onlinecite{LauchliTowers} that conformal towers for different critical models can be extracted via the entanglement entropy calculated in the ground state. Both the results of Ref.\onlinecite{LauchliTowers} and the present results show that the ground-state contains not only the central charge and the critical exponents, but essentially all the information about the critical theory.

Alternatively, a poor convergence of excited states in gapped systems can be explained by finite size of the MPS. As pointed above, the Schmidt values decays exponentially fast in the gapped system and the number of states with  weight above the machine precision is relatively small. Thus, the basis chosen for the ground-state might not be complete enough to describe the excitations. An exact MPS representation (e.g. AKLT state) and highly local Hamiltonians ( e.g. decoupled dimers discussed above) are extreme cases of restricted MPS basis. 
According to the area law the entanglement entropy scales as $S\propto N^{d-1}$ the number of non-zero Schmidt eigenvalues in one-dimensional system ($d=1$) is constant with the system size. 
in general  in critical systems the entanglement entropy has logarithmic corrections to the area law and scales with system size as $\propto \log N$. It implies that when the Schmidt values decay too fast on a small systems so the excitations cannot be extracted properly, sometimes the problem can be resolved by increasing the system size. 

In any case, what is really interesting from a practical point of view is that there is a one-to-one correspondance between eigenenergies that are completely
flat during a portion of the sweep, and exact eigenenergies of the system: In the critical system, flat energies reconstruct exactly and precisely the entire conformal tower with no energy missing, while in the gapped phases, the eigenenergies are never really flat, apart from the ground state energy (which is of course exact) and possibly the in-gap state (see Section \ref{sec:edge}). This suggests that looking for flat eigenenergies of the effective Hamiltonian during sweeps might be a very economical way of identifying excited states. As we will now demonstrate, this actually works in many different situations.


\section{Conformal towers of critical models}

\subsection{Ising model in a transverse field}
\label{sec:minmod_ising}

In this section, we continue the investigation of the critical transverse field Ising model given by the Hamiltonian of Eq.(\ref{eq:isingmod}). We concentrate on open chains with
different boundary conditions, and we use DMRG to extract the conformal towers that we compare with the predictions of boundary CFT. Apart from the case of free
boundary conditions, the model cannot be mapped on free fermions with a Jordan-Wigner transformation. 

The critical Ising model is described in the context of CFT by the minimal model defined by $(p,p^\prime)=(4,3)$\cite{Belavin:1984vu,francesco}. A brief review of the 
general properties of the minimal models is included in Appendix \ref{sec:minmid}. The central charge is given by $c=1-6(p-p^\prime)^2/pp^\prime=1/2$.
This minimal model has three operators: the identity $I$ with conformal dimension $h_{1,1}=0$, the spin operator $\sigma$ with $h_{1,2}=1/16$, and the energy density $\epsilon$ with conformal dimension $h_{2,1}=1/2$.

The finite size spectra of the critical Ising model on an open chain with different boundary conditions have been worked out by Cardy\cite{Cardy89}.
For free boundary conditions, the excitation spectrum is the superposition of the $I$ and $\epsilon$ conformal towers. The two towers appear separately
when the edge spins are fixed: the identity conformal tower $I$ is realized for $\uparrow ,\uparrow$ (and $\downarrow ,\downarrow$) boundary conditions, while the $\epsilon$ one is realized for 
$\uparrow ,\downarrow$ (and $\downarrow ,\uparrow$) boundary conditions. Finally,  
the $\sigma$ conformal tower is induced by mixed boundary conditions that fix the spin at one edge to be either $\uparrow$ or $\downarrow$ while the spin at the other edge remains free.

The multiplicities of the excited states can be read out from the characters of these conformal towers, which have first been
calculated in Ref.\onlinecite{rocha-caridi}. Their expansions up to order 8 are listed below for convenience:
\begin{widetext}
\begin{equation}
 \chi_I=q^{-1/48}\left(1+q^2+q^3+2q^4+2q^5+3q^6+3q^7+5q^8+...\right)
  \label{eq:char_I}
\end{equation}
\begin{equation}
  \chi_\epsilon=q^{1/2-1/48}\left(1+q+q^2+q^3+2q^4+2q^5+3q^6+4q^7+5q^8+...\right)
  \label{eq:char_eps}
\end{equation}
\begin{equation}
  \chi_\sigma=q^{1/16-1/48}\left(1+q+q^2+2q^3+2q^4+3q^5+4q^6+5q^7+6q^8+...\right)
  \label{eq:char_sigma}
\end{equation}

\end{widetext}

The terms inside the brackets gives the structure and the multiplicities of the excitation spectrum: a term $mq^n$ means that the $n$-th energy level has multiplicity $m$. The ground state corresponds to $n=0$. Accordingly, the scaling of the ground-state of a given tower is given by :
\begin{equation}
  E=\varepsilon_0N+\varepsilon_1+{\pi v\over N}\left[-{1\over 48}+x\right]
\label{eq:gsscaling}
  \end{equation}
where $\varepsilon_0$ and $\varepsilon_1$  are non-universal constants and $x$ is the corresponding conformal dimension: $x=h_{I}=0$ for the identity, $x=h_{\epsilon}=1/2$ for $\epsilon$ and $x=h_{\sigma}=1/16$ for $\sigma$ conformal towers. When the excitation spectrum is given by the superposition of several conformal towers, the corresponding characters are added. In that case, the $x$ that appears in the finite-size scaling of the ground state energy is equal to the smallest conformal dimension.
For free boundary conditions, since $h_I<h_\epsilon$, the ground-state belongs to the $I$ conformal tower and its energy scales according to Eq.\ref{eq:gsscaling} with $x=0$.

\begin{widetext}

\begin{figure}[h]
\centering 
\includegraphics[width=1\textwidth]{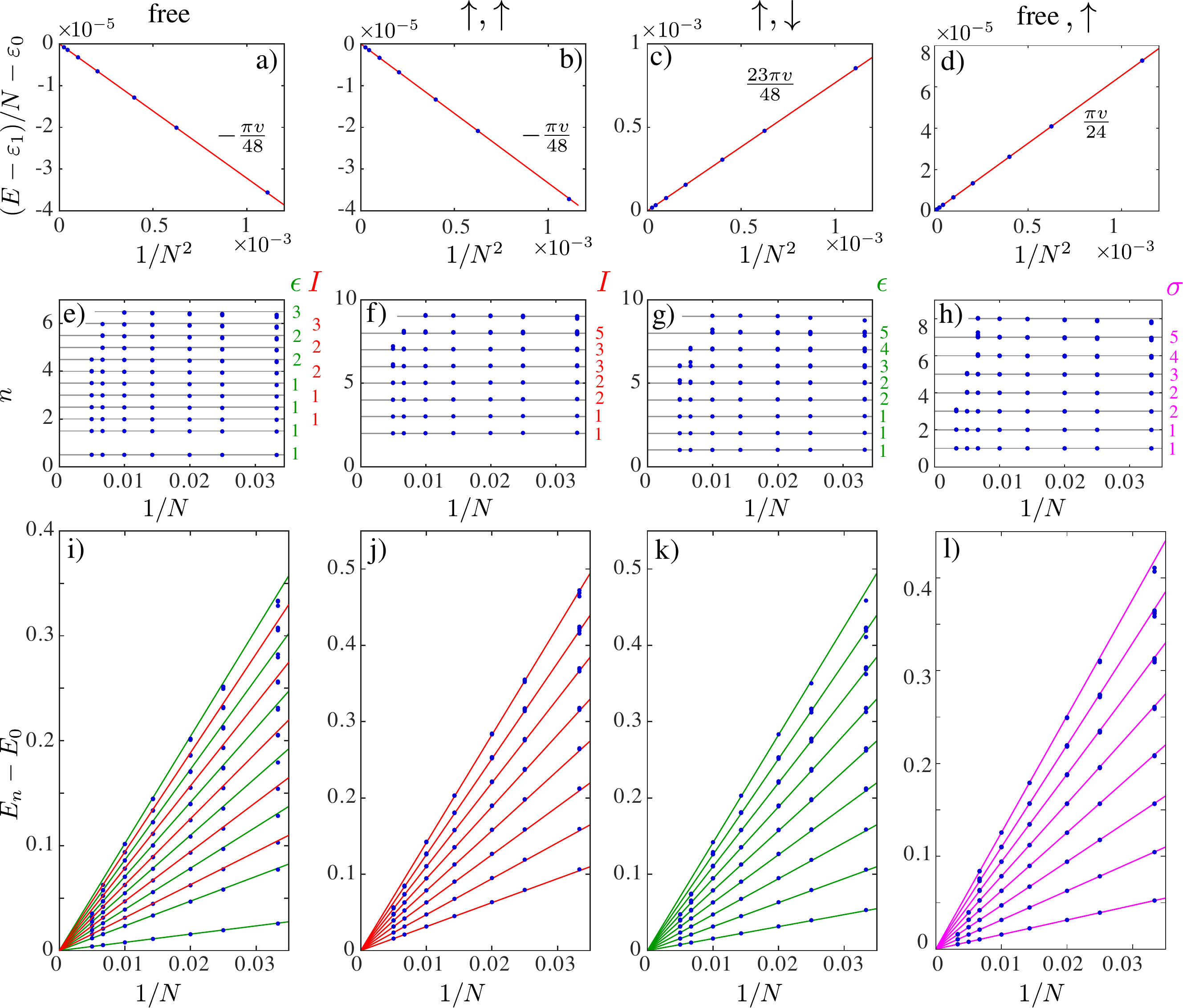}
\caption{(Color online)  Finite-size scaling of the energies of the critical Ising model in open chains with different boundary conditions: (a),(e),(i) - free at both edges; (b),(f),(j) - fixed with both edge spins pointing up; (c),(g),(k) - fixed with one edge spin pointing up and the other pointing down; (d),(h),(l) - fixed at one edge and free at the other one. (a),(b),(c) and (d) Finite-size scaling of the ground-state energy after removing the ground-state energy in the thermodynamic limit $\varepsilon_0$ and the boundary terms $\varepsilon_1$. (e), (f), (g) and (h) Conformal towers of the excitation spectra. Blue dots are the DMRG data for $n\equiv(E_n-E_0)/ (\pi vN)$ with $v=v_{Ising}=1/2$. The CFT predictions are shown with grey lines for reference. The multiplicities of the levels are indicated on the right of each tower. (i), (j), (k) and (l) Finite-size scaling of the excitation energies. Blue dots are DMRG data, red, green and magenta lines are conformal towers for the identity $I$, energy $\epsilon$ and spin $\sigma$ fields.}
\label{fig:minmod_ising}
\end{figure}

\end{widetext}

We have computed the excitation spectrum of the critical Ising model in open chains with different boundary conditions with DMRG by following many eigenvalues of the effective Hamiltonian during sweeps. The resulting finite-size spectra are shown in Fig.\ref{fig:minmod_ising}. Figures \ref{fig:minmod_ising}(a),(b),(c) and (d) show the scaling of the ground-state energy for different boundary conditions. The non-universal constants $\varepsilon_0$, $\varepsilon_1$ and the velocity $v$ are treated as fitting parameters. The values of the velocities $v_\mathrm {free}=0.491$, $v_{\uparrow\uparrow}=v_{\uparrow\downarrow}=0.509$ coincide within 2\% with the exact value $v_\mathrm{Ising}=1/2$.

Figures \ref{fig:minmod_ising}(e),(f),(g) and (h) keep track of the levels of the conformal towers obtained numerically by reporting the value of $n\equiv(E_n-E_0)/ (\pi v_{Ising}N)$. The DMRG results (dots) are in good agreement with CFT predictions (grey lines). Note that due to the absence of logarithmic corrections in the Ising model, the structure of the conformal tower is independent of the system size and can already be observed for very small system sizes ($N\approx 30-40$). The multiplicities of each level observed numerically are shown on the right of the Figures \ref{fig:minmod_ising}(e),(f),(g) and (h). They coincide with the expansion of the characters of Eq.\ref{eq:char_I}, Eq.\ref{eq:char_eps} and  Eq.\ref{eq:char_sigma}.

The finite-size scaling of the gaps is shown in Figures \ref{fig:minmod_ising}(i),(j),(k) and (l). They contain essentially the same information as the three figures above. The DMRG results for up to 20 states are marked with blue dots while red, green and magenta lines show the CFT prediction for the scaling of the $I$ and $\epsilon$ and $\sigma$ conformal towers respectively. 
The agreement is excellent. 

Previously, the three lowest levels of the conformal towers of the critical Ising model have been computed numerically by Evenbly and Vidal\cite{vidal} using scale-invariant MERA.  Our results agree with theirs, but without much numerical effort we have been able to get many more states. This 
flexibility can be useful in general to distinguish between different boundary conformal field theories. For instance, the three lowest excitation energies in the conformal towers of 
$\epsilon$ (given by  Eq.\ref{eq:char_eps}) and $\sigma$ (given by Eq.\ref{eq:char_sigma}) have the same degeneracy, and the two conformal towers only differ starting from the fourth excitation level. The conformal towers (up to twelve levels) have also been obtained from the entanglement spectrum in chains with free and periodic boundary conditions by L{\"a}uchli \cite{LauchliTowers}.


\subsection{The quantum three-state Potts model}
\label{sec:minmod_potts}

The next minimal model that we probe numerically is the three-state Potts model, which is a generalization of the transverse field Ising model to a system with a local Hilbert of dimension $d=3$. For convenience, we label single-particle states by A, B and C.
The model can be defined by the Hamiltonian\cite{rapp}:
\begin{equation}
  H_\mathrm{Potts}=-J\sum_{i=1}^{N-1}\sum_{\mu=1}^3P_i^\mu P_{i+1}^\mu-h\sum_{i=1}^NP_i,
\end{equation}
where $P_i^\mu=|\mu\rangle_{ii}\langle \mu|-1/3$ tends to project the spin at site $i$ along the $\mu$ direction while $P_i=|\lambda_0\rangle_{ii}\langle \lambda_0|-1/3$ tends to align spins along the direction $|\lambda_0\rangle_i=\sum_\mu |\mu \rangle\sqrt{3}$. The first term in the Hamiltonian plays the role of the ferromagnetic interaction, while the second one is a generalized transverse field. The model is critical and integrable for $h=J$. The critical theory is again described by a mininimal model of CFT with
$(p,p^\prime)=(6,5)$\cite{Dotsenko:1984if,Temperley:1971iq,francesco}. Its central charge is given by $c=4/5$. This minimal model has ten primary fields, with the conformal dimensions listed in Table \ref{tbl:confdim_potts}.

\begin{table}[h]
\centering
\begin{tabular}{c||c|c|c}
 & $s=1$ & $s=2$ & $s=3$\\
\hline
\hline
$r=1$ & 0 & 1/8 & 2/3 \\
 \hline
 $r=2$ &2/5 & 1/40 & 1/15\\
 \hline
 $r=3$ & 7/5 & 21/40 & \\
 \hline
 $r=4$ & 3 & 13/8 &  \\
\end{tabular}
\caption{Conformal dimension $h_{r,s}$ of the fields $\phi_{(r,s)}$ in the critical three-state Potts model. Conformal dimensions that just repeat values
realized for smaller integers are not included.}
\label{tbl:confdim_potts}
\end{table}

The small-$q$ expansion of the characters for these ten fields is provided in the Appendix \ref{sec:app_potts}. Six of them appear in the description of the operators $I$ of zero dimension, $\sigma$ of dimension $1/15$, $\epsilon$ of dimension $2/5$, and $\psi $ of dimension $2/3$. The corresponding characters are:
\begin{widetext}
\begin{equation}
  \chi_I=\chi _{1,1}+\chi _{4,1} \ \ \ \ \ \ \ \ \ \  \chi_\epsilon=\chi _{2,1}+\chi _{3,1} \ \ \ \ \ \  \ \ \ \ 
  \chi_\sigma=\chi_{\sigma^\dagger}=\chi _{2,3} \ \ \ \ \ \ \ \ \ \ 
  \chi_\psi=\chi_{\psi^\dagger} =\chi _{1,3}
\end{equation}
The small-$q$ expansion of these characters up to order 6 is given by:
\begin{eqnarray}
  \chi_I=q^{-1/30}\left(1+q^2+2q^3+3q^4+4q^5+7q^6+...\right)
  \label{eq:charpotts1}\\
    \chi_\epsilon=q^{-1/30+2/5}\left(1+2q+2q^2+4q^3+5q^4+8q^5+11q^6+...\right)
    \label{eq:charpotts2}\\
  \chi_\sigma=q^{-1/30+1/15}\left(1+q+2q^2+3q^3+5q^4+7q^5+10q^6+...\right)
   \label{eq:charpotts3}\\
  \chi_\psi=q^{-1/30+2/3}\left(1+q+2q^2+2q^3+4q^4+5q^5+8q^6+...\right)
  \label{eq:charpotts4}
\end{eqnarray}
\end{widetext}

\begin{figure}[h!]
\centering 
\includegraphics[width=0.49\textwidth]{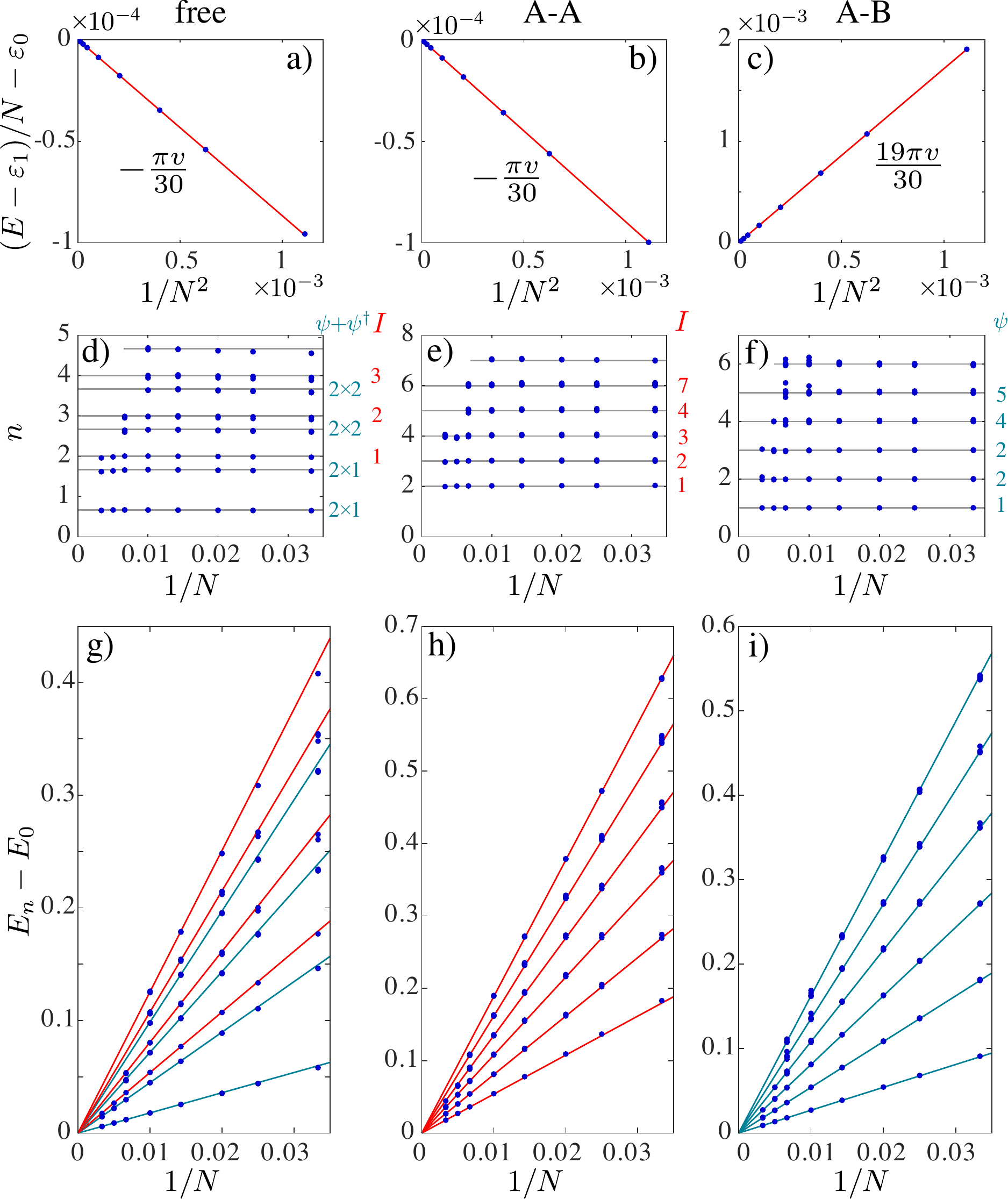}
\caption{(Color online) Finite-size scaling of the energies of the critical three-state Potts model in open chains with different boundary conditions: (a),(d),(g) - free, (b),(e),(h) - fixed with the same state at both edges, (c),(f),(i) - fixed with different states on the left and right edges. (a),(b),(c) Finite-size scaling of the ground-state energy after removing the ground-state energy in the thermodynamic limit $\varepsilon_0$ and the boundary terms $\varepsilon_1$. (d), (e), (f) Conformal towers of the excitation spectra. Blue dots are the DMRG data for $n\equiv(E_n-E_0)/ (\pi vN)$ with $v=v_\mathrm{A-A}=0.857$. The CFT predictions are shown with grey lines for reference. The multiplicities of the levels are indicated on the right of each tower. (g), (h), (i) Finite-size scaling of the excitation energies. Blue dots are DMRG data, red and blue lines are conformal towers for $I$ and $\psi$ fields respectively}
\label{fig:potts1}
\end{figure}

\begin{figure}[t!]
\centering 
\includegraphics[width=0.49\textwidth]{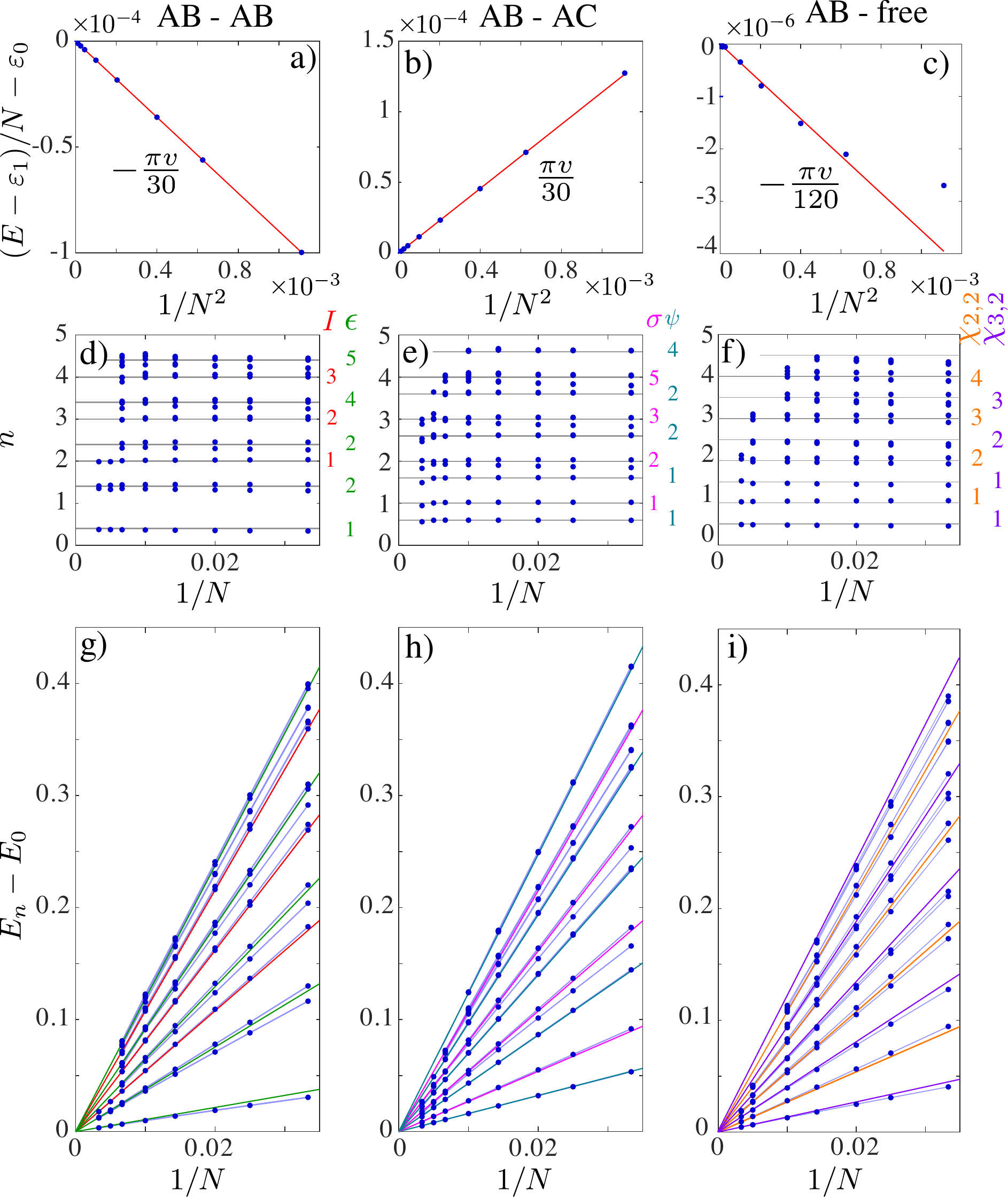}
\caption{(Color online)  Finite-size scaling of the energies of the critical three-state Potts model in open chains with partially fixed boundary conditions: (a),(d),(g) - only two states are allowed at the edges and the excluded state is the same for left and right edges; (b),(e),(h) - only two states are allowed at the edges and the excluded state is different for left and right edges; (c),(f),(i) - only two states are allowed at one edge while the other edge remains free. (a), (b) and (c) Finite-size scaling of the universal term in the ground-state energy. (d), (e) and (f) conformal towers of the excitation spectra. Blue dots are the DMRG data for $n\equiv(E_n-E_0)/ (\pi vN)$ with $v=v_{A-A}=0.857$. The CFT predictions are shown with grey lines for reference. The multiplicities of the levels are indicated on the right of each tower. (g), (h) and (i) Finite-size scaling of the excitation energies. Blue dots are DMRG data, lines of different colors correspond to different conformal towers.}
\label{fig:potts2}
\end{figure}

\begin{figure}[h!]
\centering 
\includegraphics[width=0.49\textwidth]{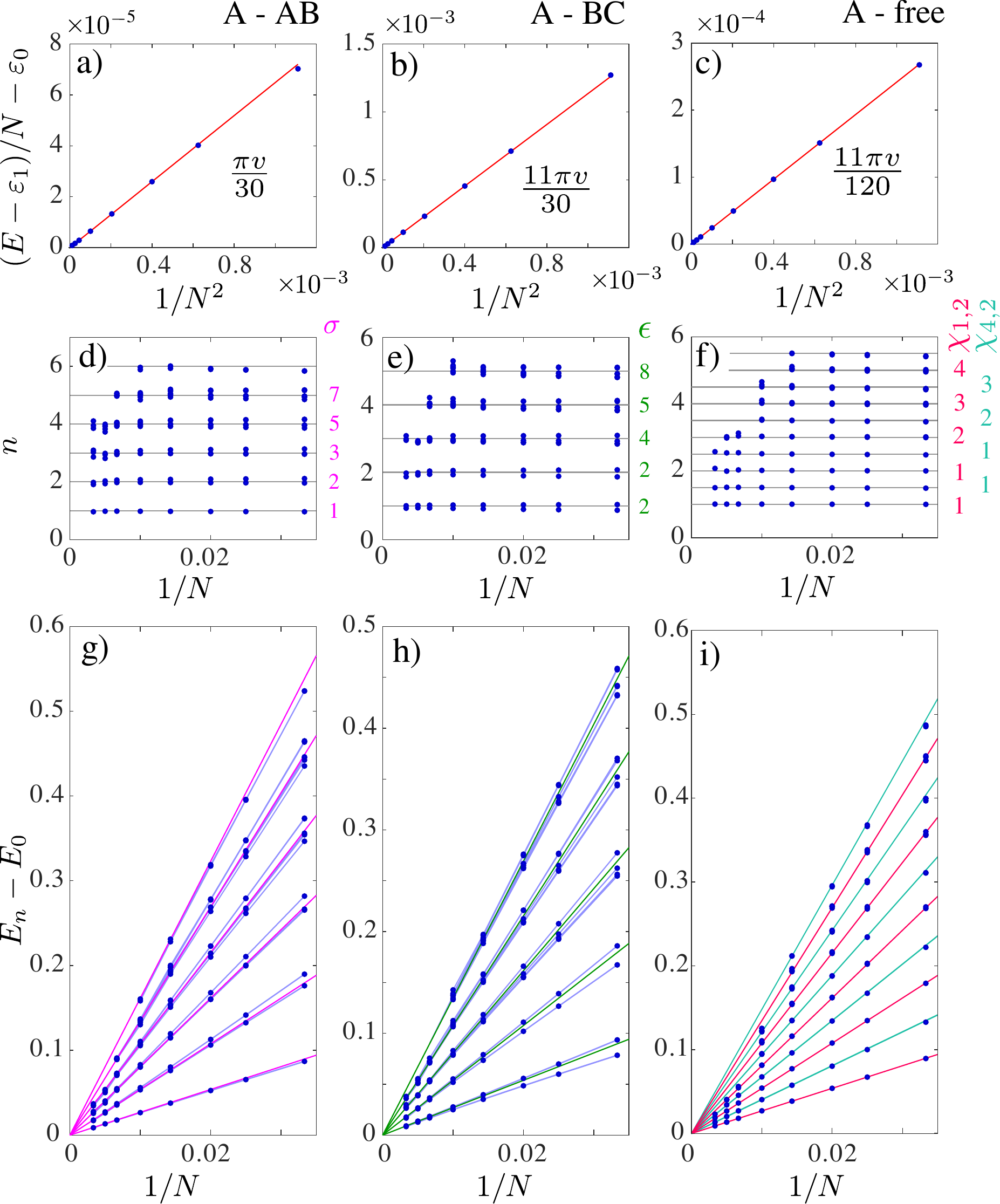}
\caption{(Color online) Finite-size scaling of the energies of the critical three-state Potts model in open chains with mixed boundary conditions: (a),(d),(g) - only two states are allowed at one edge and only one of those is allowed at the other edge; (b),(e),(h) - only two states are allowed at one edge and the third one is the only state allowed at the other edge; (c),(f),(i) - only one state is allowed at one edge while the other edge remains free. (a), (b) and (c) Finite-size scaling of the universal term in the ground-state energy. (d), (e) and (f) Conformal towers of the excitation spectra. Blue dots are the DMRG data for $n\equiv(E_n-E_0)/ (\pi vN)$ with $v=v_{A-A}=0.857$. The CFT predictions are shown with grey lines for  reference. The multiplicities of the levels are indicated on the right of each tower. (g), (h) and (i) Finite-size scaling of the excitation energies. Blue dots are DMRG data, lines of different colors correspond to different conformal towers.}
\label{fig:potts3}
\end{figure}

The appearance of different conformal towers under various applied boundary conditions was studied by Cardy\cite{Cardy89}. Our numerical results for all possible boundary
conditions obtained along the sames lines as for the Ising model are reported in Figs.\ref{fig:potts1},\ref{fig:potts2},\ref{fig:potts3}. These simulations have been performed with 
$D_\mathrm{inf}=24$, $D_\mathrm{start}=54$, and $D_\mathrm{max}=200$. 
The results agree very well with Cardy's predictions, with occasionally significant finite-size effects. In the rest of this section, we perform a detailed comparison between Cardy's predictions and our numerical results.  

In an open chain with free boundary conditions at both ends, the excitation spectrum corresponds to the superposition of three conformal towers given by $\chi_I\oplus \chi_\psi\oplus\chi_{\psi^\dagger}$. However, the characters of the field $\psi$ and its conjugate are equal: $\chi_{\psi^\dagger}=\chi_\psi$. So the spectrum looks like the superposition of the $I$ conformal tower with two copies of the
$\psi$ conformal tower. The three towers split under fixed boundary conditions. If only the local state A is allowed at both edges (boundary condition of A-A type), the excitation spectrum is given by the identity conformal tower $I$. When the allowed state is different at the edges, the boundary conditions are then A-B and A-C, and in both cases the excitation spectrum is given by the conformal tower of $\psi$, leading to two copies of this conformal tower in the spectrum. Note that the conformal towers for free and periodic boundary conditions were obtained previously from entanglement spectra by L{\"a}uchli \cite{LauchliTowers}.

 The finite-size scaling of the ground-state is given by:
\begin{equation}
  E=\epsilon_0N+\epsilon_1+{\pi v\over N}\left[-{1\over 30}+x\right],
  \label{eq:gsscalingpotts}
  \end{equation}
where $\varepsilon_0$ and $\varepsilon_1$ are non-universal constants. In the case of free and A-A boundary conditions, the ground state belongs to the conformal tower $I$ with conformal dimension $x=h_I=0$, while the spectrum of A-B (and A-C) boundary conditions belongs completely to the $\psi$ conformal tower with conformal dimension $x=h_\psi=2/3$. DMRG results on the ground-state scaling are summarized in Fig.\ref{fig:potts1}(a), (b) and (c). $\varepsilon_0$ and   $\varepsilon_1$  together with the velocity $v$ are treated as fitting parameters.
The obtained values of the velocities are $v_\mathrm{free}=0.827$, $v_\mathrm{A-A}=0.857$ and $v_\mathrm{A-B}=0.862$, in reasonable agreement with each other.

Figures Fig.\ref{fig:potts1}(d),(e) and (f) keep track of the levels of the conformal towers obtained numerically by reporting the value of  $n\equiv(E_n-E_0)/ (\pi v_{A-A}N)$. The DMRG results (dots) are in good agreement with CFT predictions (grey lines). Note that the structure of the conformal towers does not depend on the system size and the excitation spectra do not reveal finite-size corrections. The multiplicities of each level observed numerically are shown on the right of the Figures \ref{fig:minmod_ising}(d),(e) and (f) and coincide with the expansion of the characters of Eq.\ref{eq:charpotts1}, \ref{eq:charpotts2}, \ref{eq:charpotts3} and \ref{eq:charpotts4}.
Finally, the finite-size scaling of the excitation energies for different boundary conditions are provided in Fig. \ref{fig:potts1}(g),(h) and (i). The DMRG results for up to 20 states are marked with blue dots, while red and blue lines show the CFT prediction for the scaling of $I$ and $\psi$ conformal towers respectively. 

According to Cardy's prediction \cite{Cardy89}, the conformal towers of $\epsilon$ and $\sigma$ appear under partially fixed boundary conditions, when two states are allowed at the edges but not the third one. When the same pair of states are allowed at both edges (boundary conditions of AB-AB type), the energy spectrum is described by the superposition of the conformal towers of $I$ and $\epsilon$ (see Fig.\ref{fig:potts2}(d) and (g)).  The ground state scales according to Eq.\ref{eq:gsscalingpotts} with $x=h_I=0$ as shown in Fig.\ref{fig:potts2}(a). When different pairs of states are allowed at the two edges of a chain (boundary conditions of AB-AC type), the energy spectrum is a superposition of the $\sigma$ and $\psi$ conformal towers (see Fig.\ref{fig:potts2}(e) and (h)). The ground-state scales according to Eq.\ref{eq:gsscalingpotts} with $x=\min(h_\sigma,h_\psi)=h_\sigma=1/15$ as shown in Fig.\ref{fig:potts2}(b).

Surprisingly, the excitation levels that belong to $\sigma$ or $\epsilon$ towers exhibit strong finite-size effects, while towers $I$ and $\psi$ remain unaffected. It can be most clearly observed in Fig.\ref{fig:potts2}(g) and (h).
The discrepancy between the numerical data and the CFT predictions for $\sigma$ and $\epsilon$ towers is only significant for small systems $N<100$ and disappears upon approaching the thermodynamic limit, for which CFT predictions apply.

The velocity extracted from the scaling of the ground-state energy for AB-AB boundary conditions $v_{AB,AB}=0.857$ is in good agreement with previous results. By contrast, in the case of AB-AC boundary conditions the ground state energy scaling gives a velocity $v_{AB-AC}=1.05$, more than 20\% off the velocities obtained with other boundary conditions. Since the ground-state belongs to the $\sigma$ conformal tower, this discrepancy is probably a finite-size effect.

When the applied boundary condition fixes one edge and partially fixes the second one, the energy spectrum is described by only one tower. If the allowed states at the two edges are different (A-BC boundary), the whole spectrum belongs to the $\epsilon$ conformal tower, otherwise the excitation spectrum is described by the $\sigma$ conformal tower (A-AB boundary).
The ground state energy scales according to \ref{eq:gsscalingpotts} with $x=h_\epsilon=2/5$ (Fig.\ref{fig:potts3}(b)) and $x=h_\sigma=1/15$ (Fig.\ref{fig:potts3}(a)). As in the case of partially fixed boundary conditions (AB-AB and AB-AC), strong finite-size effects appear in both $\sigma$ and $\epsilon$ conformal towers (Fig.\ref{fig:potts3}(g) and (h)).

In order to see the towers for the remaining four primary fields, namely for $\phi_{(1,2)}$, $\phi_{(2,2)}$, $\phi_{(3,2)}$ and $\phi_{(4,2)}$, the boundary conditions should be fixed or partially fixed only at one edge, the spin remaining free at the other edge\cite{Cardy89}. The superposition of the towers with conformal dimensions $h_{2,2}=1/40$ and $h_{3,2}=21/40$ appears under AB-free boundary condition. The other two towers with $h_{1,2}=1/8$  and $h_{4,2}=13/8$ are superposed under A-free boundary conditions. 
In the case of AB-free boundary conditions, the ground state energy scales according to Eq. \ref{eq:gsscalingpotts} with $x=h_{2,2}=1/40$. This leads to a very small numerical prefactor for the universal term $-\pi v/120 N$ (Fig.\ref{fig:potts2} (c)) and therefore requires a much higher precision for the ground-state energy. Moreover, finite-size effects are significant in all states that belong to the $\chi_{2,2}$ tower including the ground state. These two factors lead to a poor estimate of the velocity $v_{AB-free}\approx 0.2$. Numerically, the calculated conformal towers  $\chi_{2,2}$ and $\chi_{3,2}$ are shown in Fig.\ref{fig:potts2} (f) and (i). Significant finite-size effects appear in both towers. However, by contrast to the velocity extracted from the ground-state energy scaling, the velocities extracted from the excitation energy of any of the twenty lowest excited states agree within 8\% with the reference value $v_{A-A}\approx0.857$.
This indicates that the finite-size effects partially cancel out in the difference between excited energies and the ground state energy.

If the spin is fixed only at one edge (A-free boundary condition) $x=h_{1,2}=1/8$ in the ground-state energy scaling of Eq.\ref{eq:gsscalingpotts} (Fig.\ref{fig:potts3} (c)). The calculated conformal towers match the theoretical predictions for $\chi_{1,2}$  and $\chi_{4,2}$. Note that the finite-size discrepancy does not appear in these two towers and the structure of the energy spectrum is clear starting from small systems (Fig.\ref{fig:potts3} (f) and (i)).

\begin{figure}[t]
\centering
\includegraphics[width=0.49\textwidth]{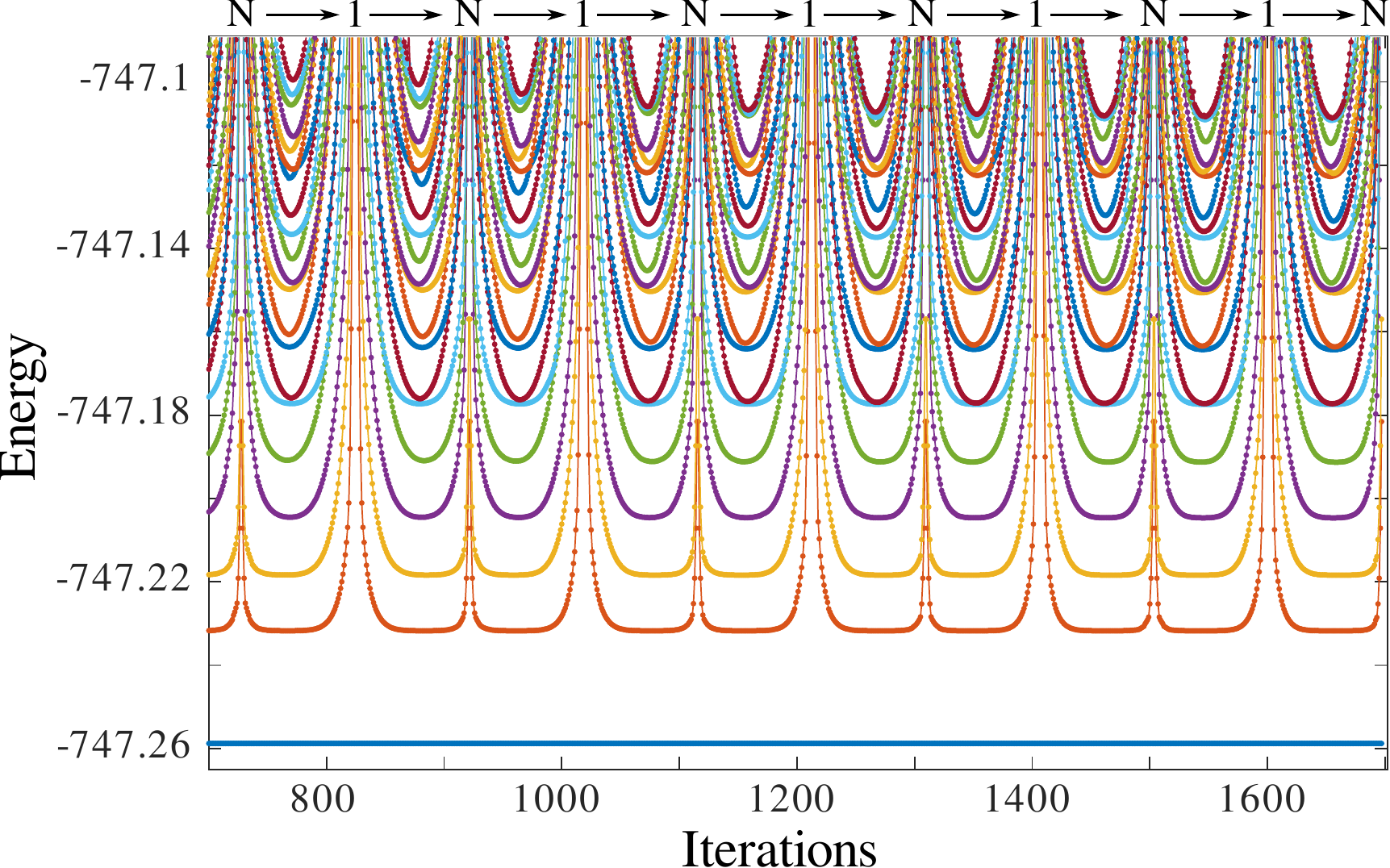}
\caption{(Color online) Energy of the ground state and of 20 low-lying excited states in the critical three-state Potts model as a function of iterations. The results are obtained for open chains with $N=100$ spins with mixed boundary A-free conditions: at the left edge (site 1) the state is fixed while at the right edge (site N) the boundary spin remains free. The flattening of the energies in the middle of the chain is taken as an indication of convergence. A periodic increase of the energy occurs close to the chain boundary and is the result of the reduced Hilbert by MPS construction. Remarkably, the boundary effect on the convergence disappears faster close to the free edge than in the vicinity of the fixed one.}
\label{fig:pot}
\end{figure}

As an example of convergence, we show in Fig.\ref{fig:pot} the 21 lowest energy levels as a function of iterations for the A-free boundary conditions. As for the Ising model, a structure
typical of a conformal tower emerges from the energy levels that are flat at the center. Remarkably, the non-symmetric boundary conditions are reflected in the convergence. The interval over which the excitation energy is poorly estimated is significantly larger close to the left fixed edge, than close to the right free one. Since only one state is selected at the left edge, the number of basis states with essentially non-zero weight in the ground state of the left block is, roughly speaking, $d$ times smaller than the number of basis vectors with non-zero weight in the case of free boundary conditions. It is thus not surprising that in the larger effective basis the excited states are better captured than in the reduced one. 

To summarize this section, keeping track of the spectrum of the effective Hamiltonian during sweeps appear as a cheap and reliable way of accessing the conformal tower of 
critical systems. This method has allowed us to confirm Cardy's predictions regarding the conformal towers of both the transverse field Ising and three-state Potts models 
under all types of boundary conditions with very high accuracy and at moderate numerical cost.


\section{Low-lying in-gap states in topological phases}
\label{sec:edge}

In addition to critical systems, we can also expect this method to work well for in-gap states, as suggested by the results we have obtained for
the transverse field Ising model at $h<h_c$, where the two-fold degenerate ground state of the thermodynamic limit shows up as two very flat quasi-degenerate 
states  (see Fig.\ref{fig:ising_compare} g)).  Let us check this by looking at a case where the degeneracy is of direct topological nature, the spin-1 Heisenberg chain\cite{haldane}:

\begin{equation}  
H=J\sum_{i=1}^{N-1} {\bf S}_i \cdot {\bf S}_{i+1},
\label{eq:heisham}
\end{equation}
where ${\bf S}_i$ are spin-1 operators. In the following, we set $J=1$.
 The topologically non-trivial nature of the Haldane phase implies the existence of spin-1/2 edge states\cite{kennedy,hagiwara}. The coupling between these edge states decays exponentially with the size of the chain, and accordingly the 
model is expected to have two low-lying states below the bulk gap, a singlet and a triplet (known as Kennedy triplet \cite{kennedy}), the singlet being below the triplet for even chains and above it for odd chains.

\begin{figure}[t]
\centering
\includegraphics[width=0.49\textwidth]{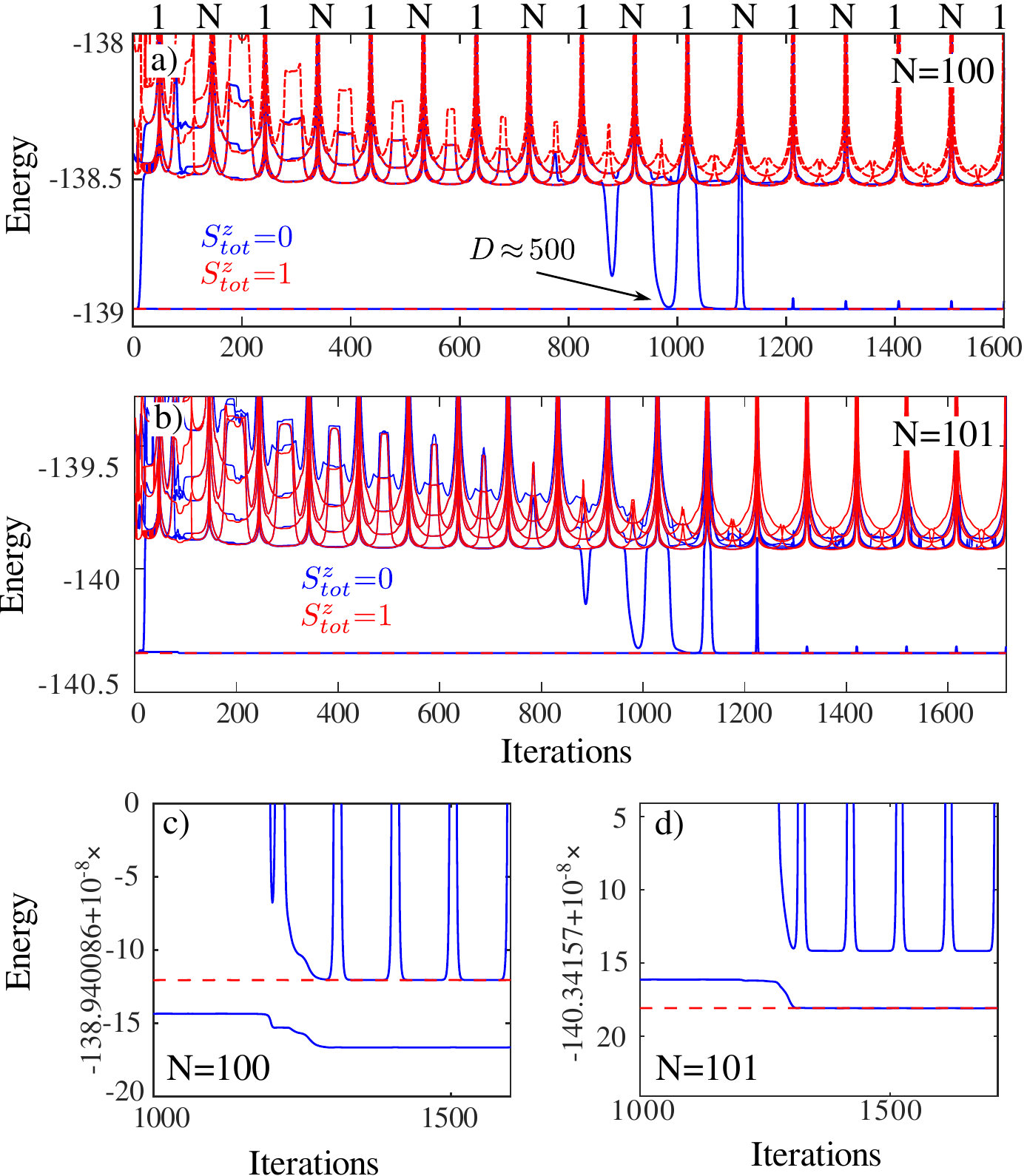}
\caption{(Color online) Ground-state and excitation energies calculated in the sectors $S^z_{tot}=0$ (blue) and $S^z_{tot}=1$ (red) in an open Haldane chain with (a) $N=100$ and (b) $N=101$ sites. (c) and (d) Enlarged parts of (a) and (b) close to the ground-state energy.}
\label{fig:haldane}
\end{figure}

By targeting several eigenvalues of the effective Hamiltonian, we have been able to extract the excitation energy of the low-lying in-gap state with sufficiently high precision, while the rest of the spectrum remains unphysical (see Fig.\ref{fig:haldane}(a),(b)). The exponentially small gap to the low-lying in-gap states requires one to keep a relatively large number of states. However, as shown in Fig. \ref{fig:haldane}(c),(d), when the number of states is large enough and the effective Hamiltonian is far enough from the edges,  the energy as a function of iterations becomes completely flat, exactly like for the excitation spectrum of critical systems.

Since for an even number of sites, the ground-state of the model is a singlet and the low-lying in-gap state is a triplet, the energy of the first excited state in the sector of zero total magnetization $S^z_\mathrm{tot}=0$ can be checked by comparing it to that of the lowest-energy state in the sector  $S^z_\mathrm{tot}=1$ in Fig. \ref{fig:haldane}(c),
and the agreement is very good as soon as the energy of that state is sufficiently flat.

Of course, the real worth of this method shows up when the first excited state cannot be calculated simply as the ground state in a different symmetry sector. For instance, in the Haldane chain with an odd number of sites, where the ground state is a triplet and the in-gap state is a singlet, this in-gap state is only accessible as the first excited state in the sector of $S^z_{tot}=0$. The results for $N=101$ are shown in Fig.\ref{fig:haldane}(b) and (d). To check the value of the gap in the case of odd chains, we have calculated it for
several system sizes, and we have compared it with the better controlled results of even chain. As expected, all these gaps fall on a single curve consistent with $\Delta \propto e^{-L/\xi}$ with $\xi\simeq 6$, a rather stringent confirmation of the accuracy of our results for odd chains. 

\begin{figure}[t]
\centering
\includegraphics[width=0.49\textwidth]{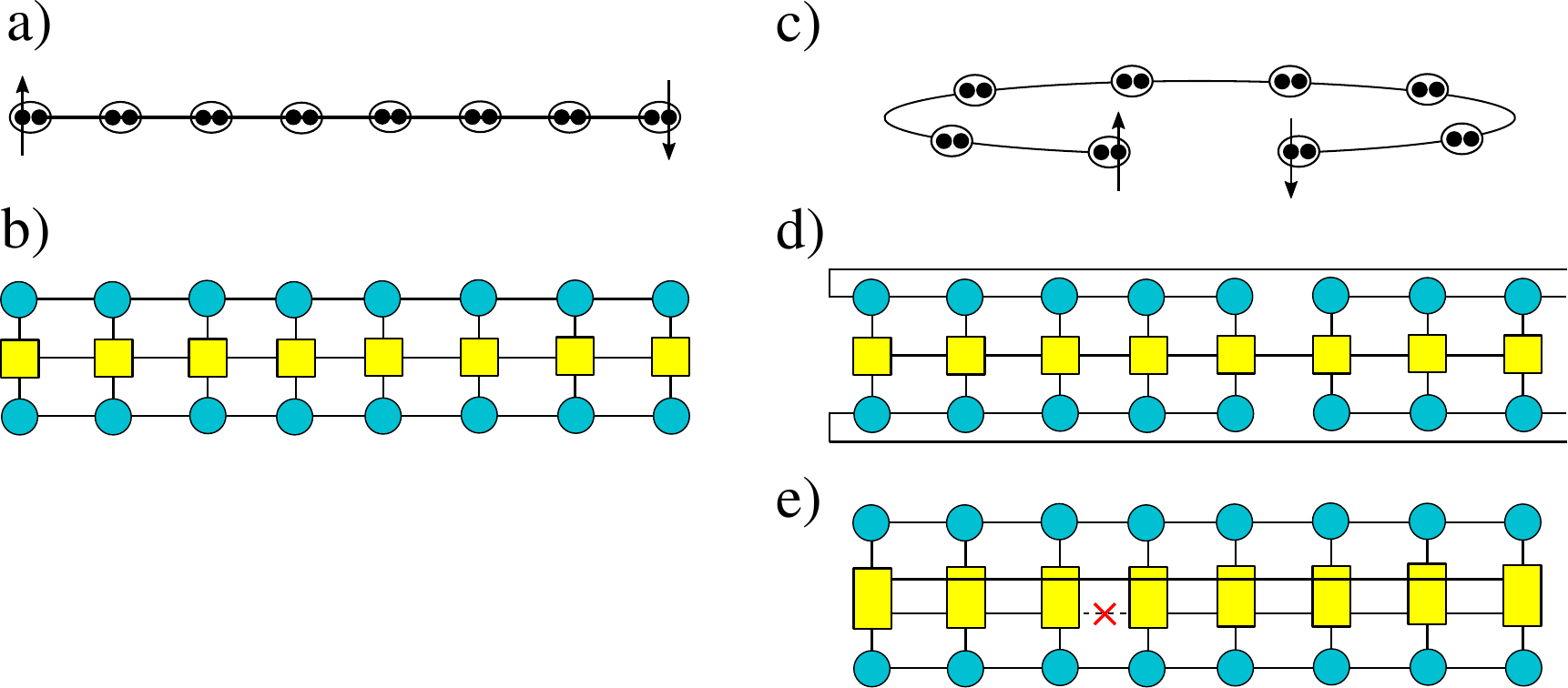}
\caption{(Color online) Transformation of an open system with edge states (left panels) into  a system where two quasi-decoupled spins are brought next to each other (right panels). (a) and (c) are VBS sketches. (b) is an MPS tensor network for (a). The tensor network for (c) can be obtained either by shift of the MPS with respect to the MPO as shown in (d), or by breaking a selected bond in the MPO for a periodic Hamiltonian as sketched in (e). From the physical point of view, (d) and (e) are equivalent. However, the implementation sketched in (e) has lower complexity.}
\label{fig:mps3}
\end{figure}

Interestingly, the idea of following the excitation spectrum during sweeps can also be used to show directly that this in-gap state is indeed the results of the coupling
between emergent spin-1/2 degrees of freedom that are localized close to the edge. The standard way of proving this consists in calculating the excited state explicitly, and in computing
observables in that state, for instance the local magnetization if this state is a triplet\cite{dmrg5,miyashita,sorensen_affleck,polizzi}. The basic idea behind the alternative we propose is that, if we were able to construct
an MPS that does not start at the edge so that the Hilbert space of the effective Hamiltonian is not small close to the edge, we might be able to follow the first excitation
close to the edge and keep track of the convergence as a function of the position of the effective Hamiltonian. 
This could be achieved by shifting the MPS with respect to the MPO as shown in Fig.\ref{fig:mps3}(d). It turns out to be more efficient to reformulate this problem in terms of a periodic Hamiltonian with one broken bond, as shown in  Fig.\ref{fig:mps3}(e). To be more specific, we include in the MPO with  open boundary conditions an additional interaction between the first and last sites represented as a long connecting line in  Fig.\ref{fig:mps3}(c). Although this increases the auxiliary (horizontal) bond dimension of the MPO from 5 to 8, the MPO on the first and last sites remains three-dimensional tensors, and not four-dimensional. Physically, there is a one-to-one correspondence between Fig.\ref{fig:mps3} (d) and (e), but the complexity is smaller for the latter case. Going from (b) to (d) the complexity increases by a factor 5, while going from (b) to (e) it increases only by a factor 1.6.

\begin{figure}[t]
\centering
\includegraphics[width=0.49\textwidth]{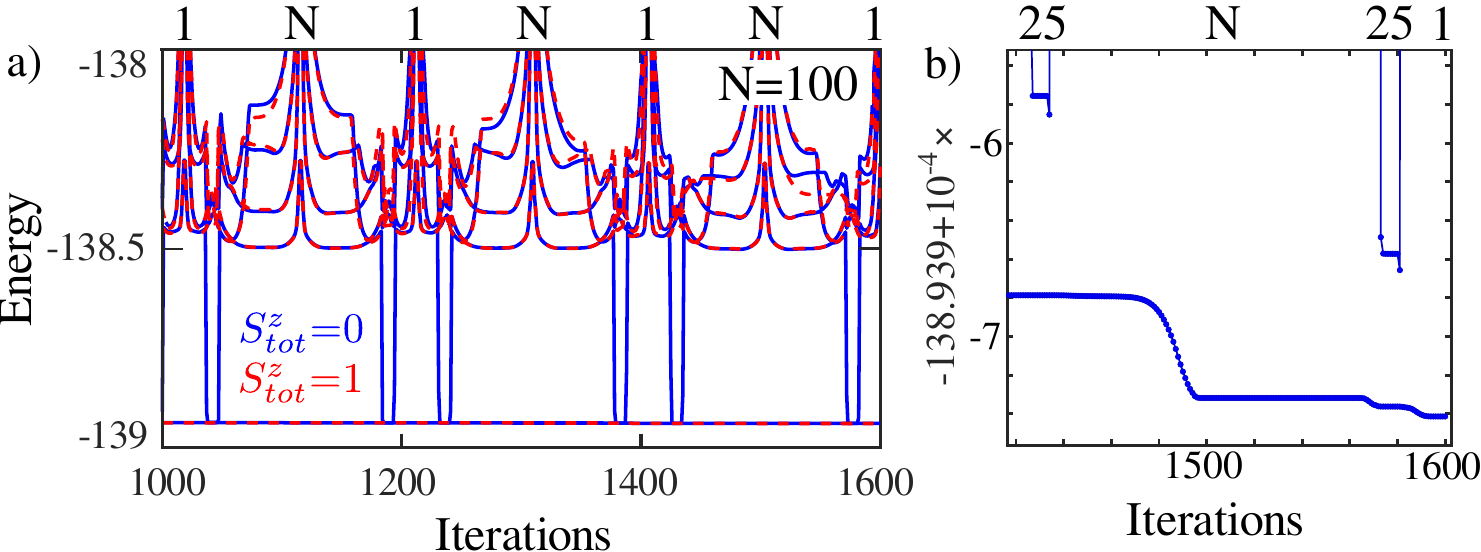}
\caption{(Color online) (a) Ground-state and excitation energies sectors of $S^z_{tot}=0$ (blue) and $S^z_{tot}=1$ (red) in a periodic Haldane chain with $N=100$ sites and one broken bond between the $25^\mathrm{th}$ and $26^\mathrm{th}$ sites. (b) Enlarged part of (a) around the ground state energy. The energy of the first excited state is flat over 7 spins in the vicinity of the broken bond.}
\label{fig:haldane_pbc}
\end{figure}

We have applied this idea to the Haldane chain with $N=100$ sites with periodic boundary conditions (i.e. with connected edges) and with a broken bond between the $25^\mathrm{th}$ and $26^\mathrm{th}$ sites,  sufficiently far from the edges, where the Hilbert space is too small, and from the middle of the chain, where the edge excitations could in principle be mixed with a bulk one. As expected, the singlet-triplet edge excitation is then seen in Fig.\ref{fig:haldane_pbc}(a) as a local excitation of the bond $(25,26)$. This local excitation can be captured immediately with the effective Hamiltonian, even if the bond dimension is small ($D<50$). 
Although, the values of the energies are less precise than in the previous case  because of the slower convergence of DMRG for periodic systems, the energy of the first excitation is completely flat over 6-7 spins in the vicinity of the broken bond. This means that {\it i)} the excitation is well localized in the vicinity of the broken bond and {\it ii)} the energy is computed exactly when the rest of the tensor network - the environment - is fixed. Thus the decrease of the energy of the in-gap state is due to the variational optimization of the environment defined by the ground state. As a confirmation, one can see on Fig.\ref{fig:haldane_pbc}(b) that the big step in the energy of the first excited state corresponds to a big step of similar magnitude in the ground state energy.

To summarize, in-gap states in topological phases appear to be accessible to this method of keeping track of the spectrum of the effective Hamiltonian during sweeps, while
the rest of the spectrum is not. In addition to giving access to a very accurate estimate of its energy, this method also gives insight into the nature of the in-gap state if 
an MPS with sufficient bond dimension close to the edge is used. Indeed, in that case the low-lying state shows up very quickly as a function of the bond dimension $D$
as a flat state around the edge. Beyond the case of edge states, for which alternative methods have long been used, this observation suggests that localized excitations
around impurities might be conveniently accessed using the simple ground state DMRG algorithm provided one keeps track of the sweeps close to the impurities. This is
the subject of the next section.  


\section{Excitations near defects and impurities}
\label{sec:local}

The effect induced by impurities in quantum chains has been studied in the past using DMRG in several ways. Some properties can be extracted from the ground-state, e.g. 
the Friedel oscillations induced by defects\cite{schmitteckert1996} or the  phase sensitivity of the ground state energy \cite{schmitteckert1998} to mention a few. However, in many practical applications, it is important to know how the energy gap changes in the presence of impurity. The problem is easy to solve when the ground-state and the excitation induced by the impurity can be distinguished by their quantum numbers\cite{mikeska_impurity,sorensen_impurity}.  When it is not the case, the energy spectra is traditionally extracted by using the mixed state approach in conventional DMRG \cite{mikeska_impurity,ng_impurities}. 

As we shall now show, localized excitations around impurities constitute another example where following the spectrum of the effective Hamiltonian as a function of 
sweeps can provide an accurate estimate for the excitation energy of gapped systems. 
In such a situation, it should be possible to write a localized excited state as the MPS of the ground-state on all sites but just a few, where the on-site tensors have to be modified to encode the excitation. Thus, it is natural to expect that the effective Hamiltonian written in the effective ground-state basis could provide not only the ground-state but also some excitation energies when optimized in the vicinity of the localized impurity. To be concrete, 
let us discuss this problem in the context of the explicitly dimerized spin-1/2 Heisenberg model with alternating coupling constants:
\begin{multline}
\label{eq:heisham}
  H=J_\mathrm{even}\sum_{i=1}^{N/2-1}{\bf S}_{2i}\cdot{\bf S}_{2i+1}+J_\mathrm{odd}\sum_{i=2}^{N/2-1}{\bf S}_{2i-1}\cdot{\bf S}_{2i}\\
  +J_{1}{\bf S}_{1}\cdot{\bf S}_{2}+J_{N}{\bf S}_{N-1}\cdot{\bf S}_{N}
\end{multline}
with $J_\mathrm{odd}=1$ and $J_\mathrm{even}=0.1$. We will concentrate on even chains for clarity. Local bond impurities are imposed at the edges by changing the coupling constants $J_1$ and $J_{N}$ on the first and last bonds. In the absence of impurities ($J_1=J_N=J_\mathrm{odd}$), the ground state of of the Hamiltonian (\ref{eq:heisham}) approximately consists of spin-$1/2$ singlets located on every strong (odd) bond. The system has a finite gap of order $J$ to the first triplet excitation. By changing locally the coupling of a particular odd bond one can control the energy needed to excite the selected bond to a triplet state. For convenience, let us introduce two different edge impurities by reducing the coupling of the first and the last bonds to $J_1=0.3$ and $J_{N}=0.8$, and let us consider a system of $N=28$ sites with open boundary conditions, 
for which exact diagonalizations can be performed, the goal being to benchmark our method. 

\begin{figure}[h!]
\centering
\includegraphics[width=0.47\textwidth]{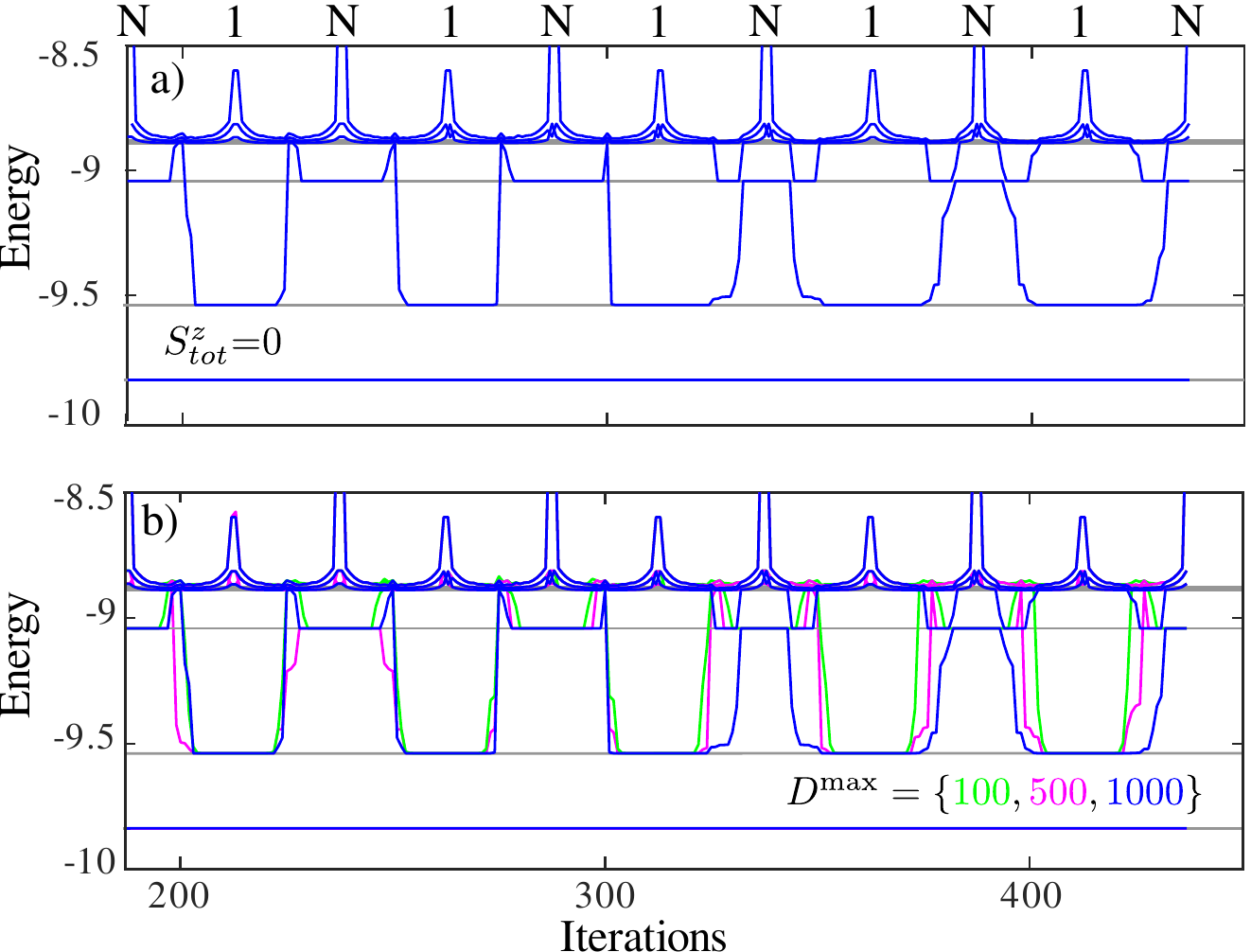}
\caption{(Color online) Local excitation energy in the dimerized Heisenberg chain with bond impurities. (a) Energy spectrum as a function of the number of iterations for the alternating Heisenberg chain with $J_\mathrm{odd}=1$ and $J_\mathrm{even}=0.1$ and weak first ($J_1=0.3$) and last ($J_N=0.8$) bonds for $N=28$ in the sector of $S^z_{tot}=0$. Results obtained with exact diagonalizations are shown with grey lines for reference. The number of states increases during the warm-up and the first five sweeps up to $D^{\max}=1000$ and is kept constant over the last three sweeps. b) Energy spectrum in the sector $S^z_{tot}=0$ for different numbers of kept states $D^{\max}=100$ (green), $500$ (magenta) and $1000$ (blue) with the same growing procedure as in (a). 
}
\label{fig:heisenberg}
\end{figure}

Using DMRG, we have calculated the ground-state energy and the energy of a few low-lying excitations in the sector $S^z_{tot}=0$. Figure \ref{fig:heisenberg} (a) shows the  energies obtained at each iteration in variational MPS by calculating several eigenvalues of the effective Hamiltonian, together with the 
exact spectrum (grey lines). The ground state has an energy of about $E_0\approx-9.838$. The spectrum computed in the middle of the chain clearly misses the first
two excitations. However, by tracking the energy as a function of DMRG iterations one can see well-converged energy levels $E_1\approx-9.539$ when the Hamiltonian is diagonalized close to the site 1 and $E_2\approx-9.042$ when the diagonalization is around site $N$. The DMRG values for these energies agree within $10^{-12}$ with 
the exact diagonalization results. Besides, the
excitation energies $E_1-E_0\approx 0.299$ and $E_2-E_0\approx 0.796$ agree with the rough estimates $\Delta_1=J_1$ and $\Delta_N=J_N$ for the singlet-triplet excitations in the case of decoupled dimers. So their interpretation as local excitations is clear.
The energies of the higher excited states around $E_{3,4}\approx-8.9$ have local minima close to the middle of the chain. They correspond to the bulk gap and agree with 
exact diagonalizations.

\begin{figure}[h!]
\centering
\includegraphics[width=0.47\textwidth]{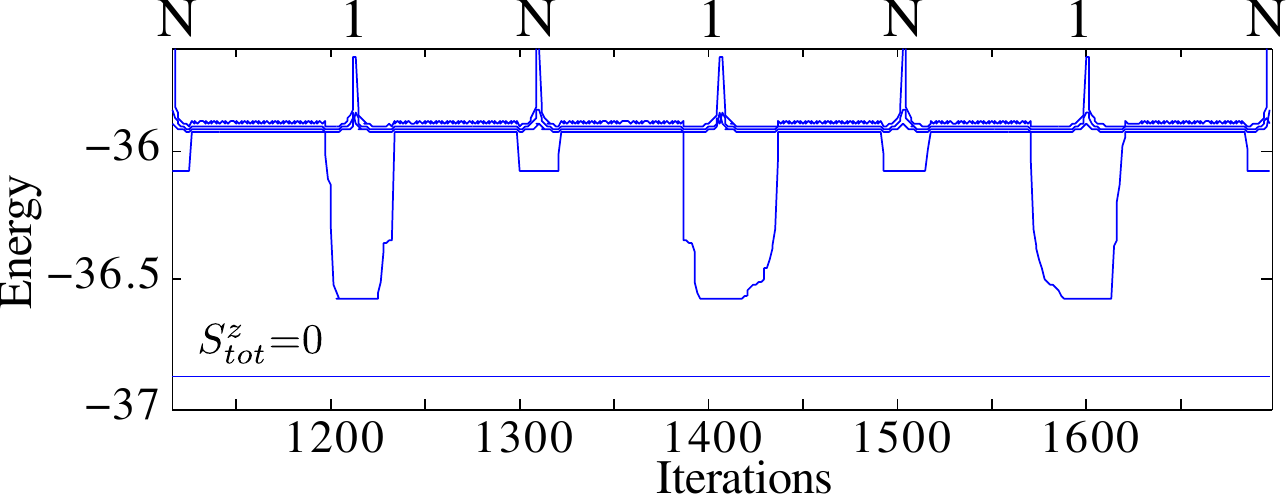}
\caption{(Color online) Energy spectrum as a function of iteration for the alternating Heisenberg chain with $J_\mathrm{odd}=1$ and $J_\mathrm{even}=0.1$ and weak first ($J_1=0.3$) and last ($J_N=0.8$) bonds for $N=100$ in the sector of $S^z_{tot}=0$. The energy of the two localized excited state only appears in the narrow window close to the edges.}
\label{fig:heisenberg_large}
\end{figure}

In these simulations, as usual, we have increased the number of kept states linearly over the first five sweeps, and then we have performed three more sweeps with $D=D^{\max}$. The results for different numbers of states are shown in Fig. \ref{fig:heisenberg}(b).
Interestingly, in most cases except the two last sweeps with $D^{\max}=1000$, there are small intervals where the first excited state in the $S^z_{tot}=0$ sector converges to the bulk excitation $E_3>E_1,E_2$ while changing between $E_1$ and $E_2$. This happens when the number of kept basis vectors for the left and right environment tensors are not sufficiently large to see the edge excitations. However, the environment tensors converge well to the ground state, and the bulk spectrum is seen properly. This is what we expect also for much larger systems - the localized edge excitations will only be seen in the vicinity of the corresponding edges, as shown in Fig.\ref{fig:heisenberg_large}. 

To summarize, although the excitation spectrum calculated in the middle of the chain is not physical in the sense that it misses the first two excitations, one can identify these
excitations as flat modes around the edges by keeping track of the energy as a function of iterations, and their energy is obtained with very high accuracy. This time, the main
reason of the success of the method is that, for localized excitations, a big portion of the system is in the same state as it is in the ground-state. In other words, the triplet excitation localized at the right edges implies that the left environment tensor is essentially the same (exactly the same for completely localized excitations) as that of the ground-state. 
On top of giving access to accurate estimates of the energies, the method also has the advantage of distinguishing the excitation energies of different impurities by the position
at which flat modes occur during the sweeps. This can be used to distinguish localized or quasi-localized edge excitations from the bulk spectrum in more complicated critical systems\cite{J1J2J3_long}.


\section{Accuracy of the wave-functions}
\label{sec:accuracy}

In this section we take a closer look at the MPS wave-functions of the excited states, and not just at their energy. 
The aim of this study is twofold: First, to check that the resulting MPS correspond to true eigenstates of the full Hamiltonian; and second, to determine the accuracy of these wave-functions.

\begin{figure}[h!]
\centering
\includegraphics[width=0.47\textwidth]{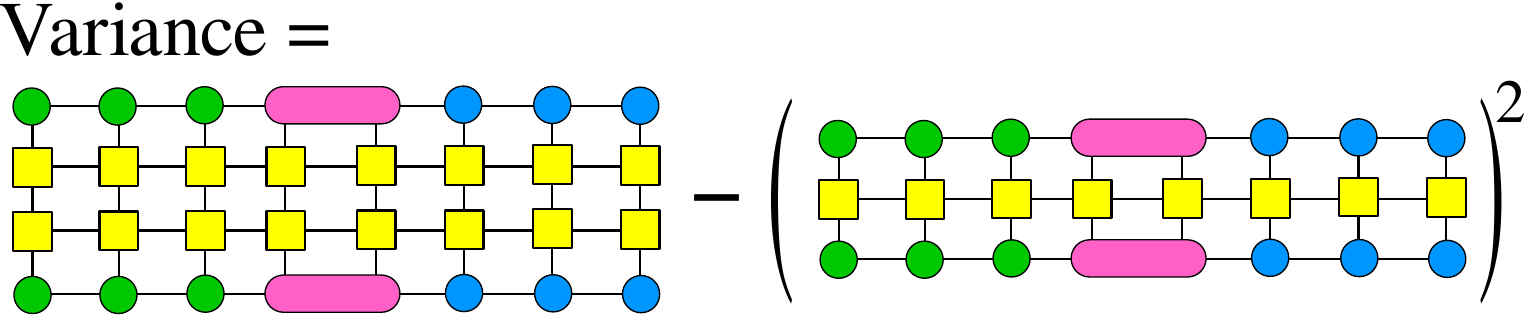}
\caption{(Color online) Graphical representation of the tensor network that corresponds to the variance: $\langle \Psi_j|\hat{H}^2|\Psi_j\rangle-\langle \Psi_j|\hat{H}|\Psi_j\rangle^2$. Green and blue boxes correspond to the left and right normalized on-site tensors of the ground state. Yellow boxes denote the on-site MPO. The operator $\hat{H}^2$ is obtained by doubling the line of the MPO. The magenta box corresponds to the various eigenstates of the effective Hamiltonian. The term in brackets is equal to a real number (energy). Thus taking square of this network is a trivial algebraic operation.}
\label{fig:variance_tn}
\end{figure}

In order to quantify the accuracy of the MPS for some state $|\Psi_j\rangle$, we compute the variance of the energy of this state: $\langle \Psi_j|\hat{H}^2|\Psi_j\rangle-\langle \Psi_j|\hat{H}|\Psi_j\rangle^2$. The corresponding tensor network structure is shown in Fig.\ref{fig:variance_tn}. 
Different excited states of the system correspond to different eigenstates of the effective Hamiltonian (magenta box). Each of these eigenstates are contracted with the tensors that belong to the left (green) and right (blue) environments and are defined by the ground state only.

\begin{figure}[h!]
\centering
\includegraphics[width=0.47\textwidth]{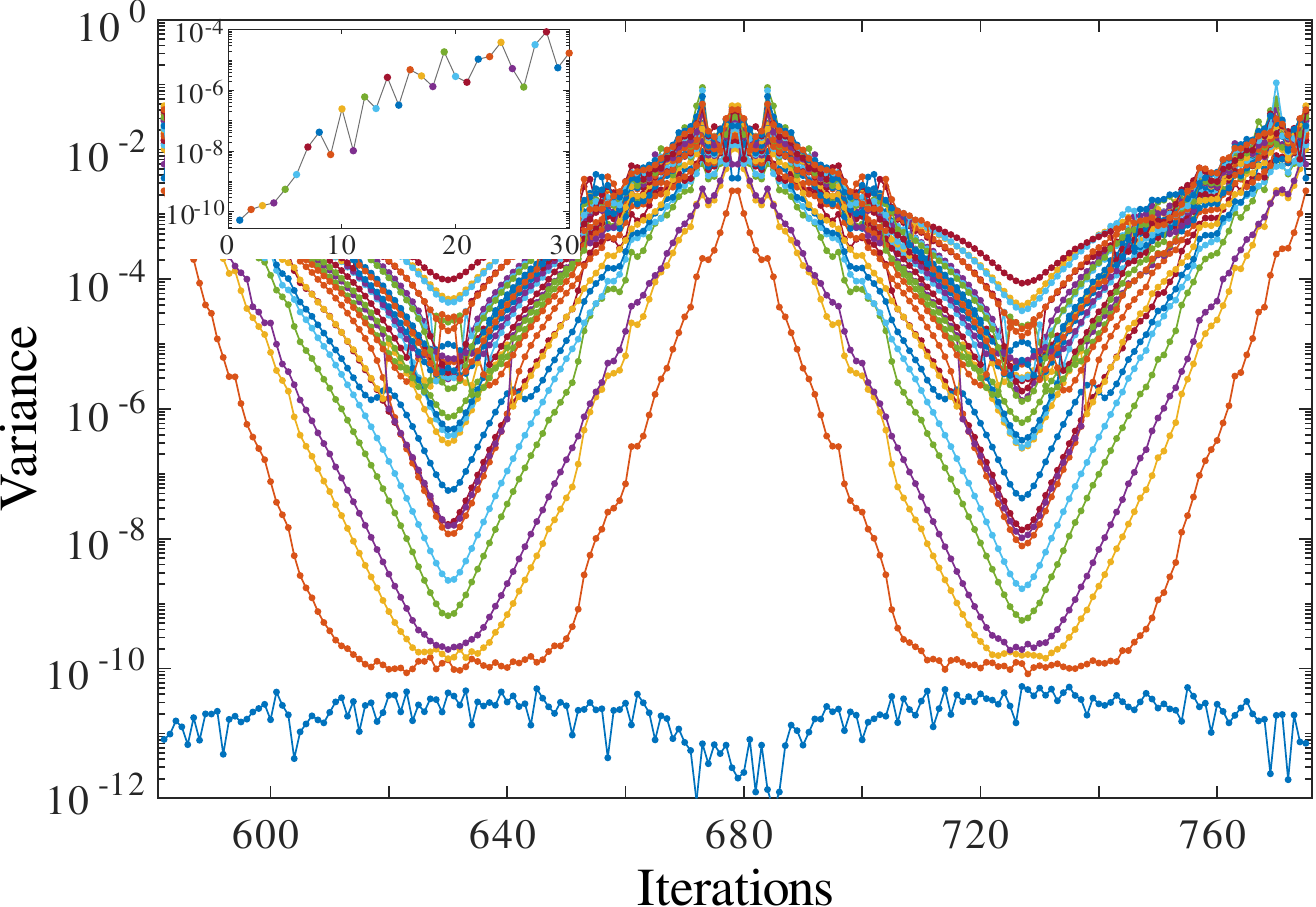}
\caption{(Color online) Variance of the Hamiltonian in the first 30 low-energy MPS states for the critical transverse field Ising model as a function of iterations during one sweep (right-to-left and left-to-right). The energy spectrum for this model is shown in Fig.\ref{fig:ising_conv}. Inset: Variance in the middle of the chain as a function of the sequential number of the excited states. }
\label{fig:ising_variance}
\end{figure}

We start by presenting results for the critical transverse field Ising model. The spectrum of the system with $N=100$ sites and free boundary conditions has been presented above in Fig.\ref{fig:ising_conv}. The variance of the resulting MPS for different excited states as a function of iterations is shown . 
As expected for bulk excitations, the variance is minimal at the center of the chain, while the accuracy of the MPS states close to the edges is much lower. 
Note that the variance in the middle of the chain of all the 30 states shown in Fig.\ref{fig:ising_variance} is below $10^{-4}$. Moreover, the six lowest states have a variance smaller that $10^{-8}$. Interestingly, the variance is a non-monotonous function of the sequential number of the excited states (see inset of  Fig.\ref{fig:ising_variance}). This signals that some of the excited states converge more slowly than states of higher energy. This also can be concluded from Fig.\ref{fig:ising_conv}, where some energy levels are essentially flat as a function of iterations, while energy levels below might still exhibit a significant curvature.

\begin{figure}[h!]
\centering
\includegraphics[width=0.47\textwidth]{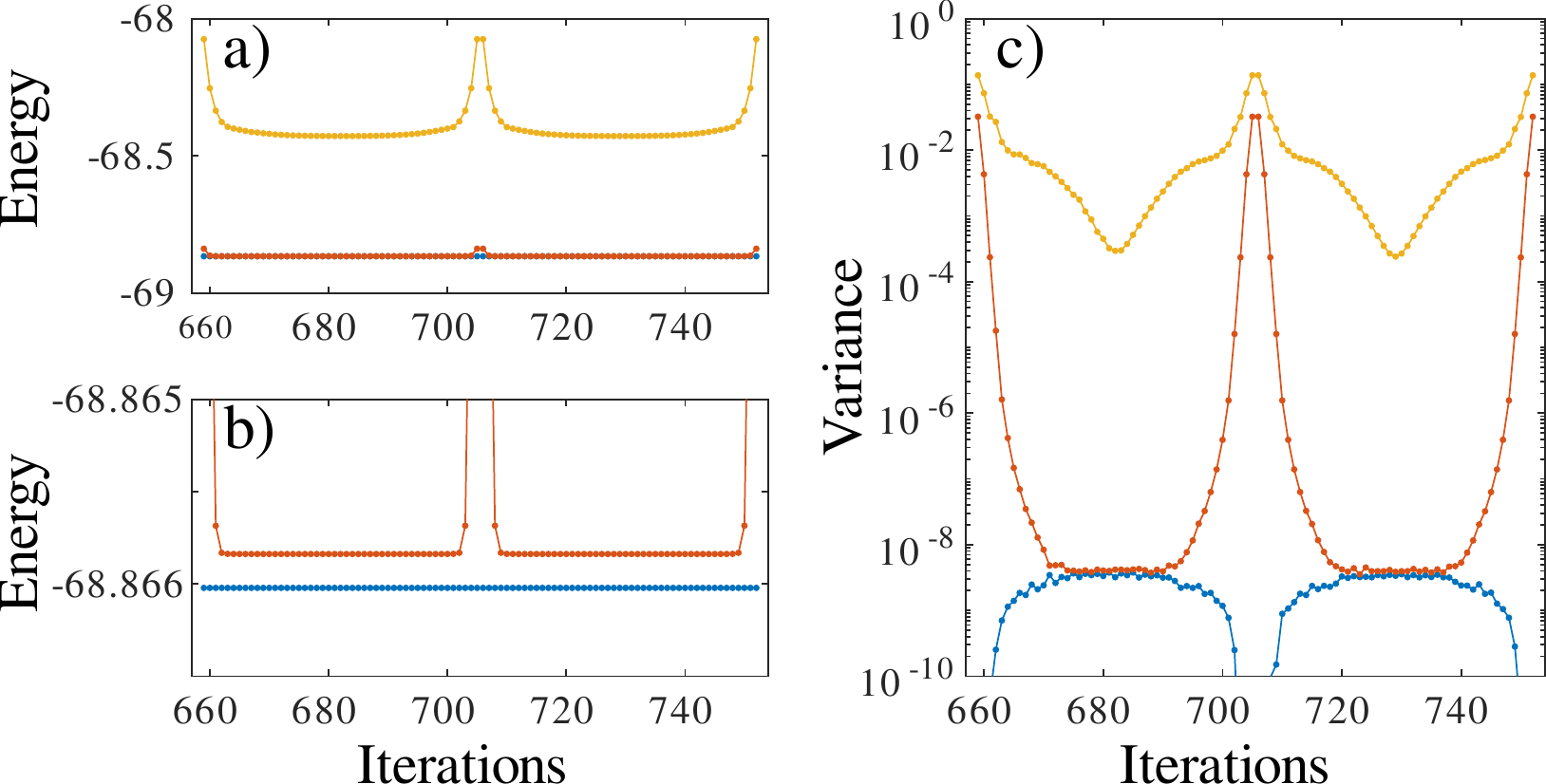}
\caption{(Color online) Accuracy of the MPS for the ground state (blue), low-lying in-gap state (red) and the bulk excitation (yellow) of a Haldane chain with $N=50$ sites. a) Energy of the three lowest states as a function of iterations during one sweep, with 800 states kept. b) Enlarged part of a). c) Variance of the Hamiltonian for these three states as a function of iterations. The MPS representation of the ground state and of the in-gap state are of the same order of accuracy. The variance for the bulk excitation is much higher. }
\label{fig:haldane_variance}
\end{figure}

As the next example we study the variance of the Hamiltonian calculated for a couple of excited states of a Haldane chain with $N=50$ sites and in the symmetry sector $S^z_{tot}=0$. The spectrum of the Haldane chain has been discussed in details in section \ref{sec:edge}. As shown in Fig.\ref{fig:haldane_variance} a) and b) the low-lying in-gap state is essentially flat as a function of iterations, while the energy of the bulk excitation has a noticeable curvature, even though the local minimum does not change much with the sweeps.  The variance of these three states is shown in Fig.\ref{fig:haldane_variance} c). The in-gap state is determined with the same order of accuracy as the ground-state, therefore both MPS representations can be used to extract the observables and characterize the properties of these states.
By contrast, the variance of the bulk excitation is of the order of $10^{-4}$. So this MPS should only be considered as a reasonable but not very accurate approximation to the true MPS for this excitation.

To summarize, if the energy of an excited state is very flat over a larger portion of the sweep, the resulting MPS representation is a true eigenstate of the total Hamiltonian with very good accuracy. In that case, the MPS representation can be used to characterize further the properties of the selected excited state by computing the correlation functions, local observables, structure factor, etc. By contrast, when the energy as a function of iterations is not flat, but its local minimum is well converged with the sweeps, the corresponding MPS is only a rough approximation of a true eigenstate of the Hamiltonian whose accuracy can be assessed by computing the variance of the energy. To use such an MPS to calculate observables would require a careful estimate of the error bars.


\section{Conclusions}
\label{sec:conclusion}

The main message of this paper relies on the empirical observation that, in certain situations, the spectrum of the effective Hamiltonian used in updating the tensors to get the 
ground state in the MPS version of DMRG contains extremely valuable information about the excited states as well. More precisely, if one keeps track of the spectrum of
the effective Hamiltonian during sweeps, energies that are completely flat during part of the sweeps, i.e. which are identical up to extremely good accuracy for a range of
neighboring effective Hamiltonians, are, in all cases we have looked at, very accurate estimates of true excitation energies. We have observed this in three different situations:

- For critical systems, such as the transverse field Ising and 3-state Potts models at their critical points. In that case, the low-lying states of the conformal tower appears as very flat
levels around the center of the chain. We have been able using this approach and for moderate computational cost to calculate the conformal towers with all types of 
boundary conditions, and thus to check the predictions of boundary conformal field theory for all primary fields. 

- For gapped systems with in-gap states due to the emergence of localized degrees of freedom at the edges. In that case, in-gap states appear as a flat level at the center
of the chain if the MPS starts at one edge and ends at the other edge, as in the usual implementation, or centered around the edge if the MPS is shifted with respect to the chain
(or equivalently if periodic boundary conditions are used, and the edge is described as a missing bond in the bulk of the chain).

- For gapped systems with defects or impurities. In that case, localized excitations appear as flat levels close to the impurities, while they can be missed altogether at the center of the system, where in general DMRG is most accurate.

By contrast, bulk excitations of gapped systems do not appear as flat levels for accessible values of the bond dimension (of course they would for large enough bond dimensions)
and more sophisticated approaches cannot be bypassed in that case. 

One interesting aspect of this method, which consists in keeping track of the spectrum of the effective Hamiltonian during sweeps, is that it is self-contained. The spectrum of
the effective Hamiltonian can always be calculated, but without keeping track of the spectrum during sweeps, it is impossible to know which eigenenergies (if any) are faithful
ones. By keeping track of the spectrum during sweeps, one can immediately spots flat levels, hence faithful eigenstates, and discard them from eigenenergies of the
effective Hamiltonian which do not correspond to exact energies. This is somehow similar to the Lanczos algorithm\cite{lanczos}, where eigenenergies can only be trusted when they 
have converged upon increasing the number or iterations. In some sense, the effective Hamiltonian corresponds to the original Hamiltonian written in a truncated basis,
a little bit like in Lanczos, where the Hamiltonian is written in a truncated Krylov basis\cite{krylov}. The analogy stops there however. In the present approach, the convergence can
be checked by comparing the energies at different sites during a sweep without increasing the size of the truncated Hilbert space. Besides, the basis has a real space
dependence that allows one to access information on the nature of the excited states. 

In all cases, we have come up with intuitive arguments to explain the presence of "exact" excitation energies in the effective Hamiltonian of the ground state, but one should keep in mind that the most solid piece of evidence clearly comes from the excellent agreement with exact results whenever available. If would thus be very interesting to see if one can put this method on a more formal basis. This clearly goes beyond the scope of the present paper.

\section{Acknowledgments} 

We are indebted to Fabien Alet and Gr\'egoire Misguich for very stimulating questions that led to the investigation reported in this paper. We would also
like to thank Salvatore Manmana and  Ulrich Schollw\"{o}ck  for insightful discussions about the standard methods to access eigenstates with DMRG. Finally, we would
like to thank Ian Affleck for teaching us the basics of boundary conformal field theory in the context of another project.
 This work has been supported by the Swiss National Science Foundation.

\appendix
\section{Minimal model}
\label{sec:minmid}

A model is called minimal if the corresponding CFT contains a finite number of local fields with well-defined scaling behavior.
The minimal models can be labeled by two positive integers $(p,p^\prime)$ that reflect the periodicity properties of the conformal dimensions:
\begin{equation}
  h_{r,s}=h_{r+p^\prime,s+p}.
\end{equation}
The central charge of the critical theory can be expressed in terms of these integers as \cite{francesco}:
\begin{equation}
\label{eq:cc_minmod}
  c=1-6\frac{(p-p^\prime)^2}{pp^\prime}
\end{equation}
For minimal models the conformal dimension $h$ is given by the Kac formula\cite{Kac1979,francesco}:
\begin{equation}
\label{eq:scaldime_minmod}
  h_{r,s}=\frac{(pr-p^\prime s)^2-(p-p^\prime)^2}{4pp^\prime},
\end{equation}
where the pair of integers $(r,s)$ label the various conformal dimensions and range in the intervals $1\leq r\leq p^\prime-1$ and $1\leq s\leq p-1$. The conformal dimension obeys the following symmetry property:
\begin{equation}
  h_{p^\prime-r,p-s}=h_{r,s}
\label{eq:minmod_sym}
\end{equation}

A minimal model is unitary if and only if $|p^\prime-p|=1$ \cite{PhysRevLett.52.1575,francesco}. In that case, the minimal conformal dimension is $h_{1,1}=0$ and it corresponds to the identity primary conformal field $\phi_{(1,1)}=I$. Assuming without loss of generality that $p>p^\prime$, then for unitary minimal models the central charge of Eq.\ref{eq:cc_minmod} can be rewritten as
\begin{equation}
\label{eq:cc_minmod_unitary}
c=1-\frac{6}{p(p-1)}.
\end{equation}
The first non-trivial $(c>0)$ unitary model is labeled by $(4,3)$ and corresponds to the critical Ising model in a transverse field\cite{Belavin:1984vu,francesco}. The following pairs $(5,4)$ and $(6,5)$ label the tri-critical Ising model \cite{Friedan:1984rv,francesco} and the three-state Potts model \cite{Dotsenko:1984if,Temperley:1971iq,francesco} respectively. 

One of the fundamental characteristics of a conformal field theory is its central charge, as first realized in Ref.\onlinecite{Belavin:1984vu}. In practice, the universality class of a critical theory is usually determined by extracting the central charge from numerical data. Although it is not always feasible to deduce the critical theory from the central charge in a unique way, the number of candidates for possible CFTs is reduced to just a few. The selection among them can often be based on simple physical intuition. In addition, the critical exponents can often be extracted from the scaling of some physical operators (on-site magnetization, correlations etc.).
Moreover, in the case of conformally invariant boundary conditions, the excitation spectrum forms a so-called conformal tower. Various boundary conditions correspond to primary fields with different conformal dimensions, resulting for the spectrum of a model with a given set of boundary conditions in one or the other conformal tower, or in the superposition of some of them. 

In the minimal model labeled by $(p,p^\prime)$, the irreducible characters are given by

\begin{equation}
\label{eq:characters}
  \chi_{(r,s)}(q)=K_{r,s}^{(p,p^\prime)}(q)-K_{r,-s}^{(p,p^\prime)}(q),
\end{equation}
where
\begin{equation}
  K_{r,s}^{(p,p^\prime)}(q)=\frac{q^{-1/24}}{\varphi (q)}\sum_{n\in \mathbb Z}q^{(2pp^\prime n+pr-p^\prime s)^2/4pp^\prime},
\end{equation}
and $\varphi (q)$ is the Euler function:
\begin{equation}
  \frac{1}{\varphi (q)}=\prod_{n=1}^\infty \frac{1}{1-q^n}
\end{equation}
The structure of the excitation spectra for a particular CFT can be deduced from the small-$q$ expansion of the characters in Eq.\ref{eq:characters}.

\section{Three-state Potts minimal model}
\label{sec:app_potts}

The small-$q$ expansions of the characters for the ten primary fields of the theory are given by:

\begin{widetext}

\begin{eqnarray}
    \chi_{(1,1)}(q)=q^{-1/30}\left(1+q^2+q^3+2q^4+2q^5+4q^6+...\right)\\
    \chi_{(2,1)}(q)=q^{-1/30+2/5}\left(1+q+q^2+2q^3+3q^4+4q^5+6q^6+...\right)\\
    \chi_{(3,1)}(q)=q^{-1/30+7/5}\left(1+q+2q^2+2q^3+4q^4+5q^5+8q^6+...\right)\\
    \chi_{(4,1)}(q)=q^{-1/30+3}\left(1+q+2q^2+3q^3+4q^4+5q^5+8q^6+...\right)\\
    \chi_{(1,2)}(q)=q^{-1/30+1/8}\left(1+q+q^2+2q^3+3q^4+4q^5+6q^6+...\right)\\
    \chi_{(2,2)}(q)=q^{-1/30+1/40}\left(1+q+2q^2+3q^3+4q^4+6q^5+9q^6+...\right)\\
    \chi_{(3,2)}(q)=q^{-1/30+21/40}\left(1+q+2q^2+3q^3+5q^4+7q^5+10q^6+...\right)\\
    \chi_{(4,2)}(q)=q^{-1/30+13/8}\left(1+q+2q^2+3q^3+4q^4+6q^5+9q^6+...\right)\\
    \chi_{(1,3)}(q)=q^{-1/30+2/3}\left(1+q+2q^2+2q^3+4q^4+5q^5+8q^6+...\right)\\
    \chi_{(2,3)}(q)=q^{-1/30+1/15}\left(1+q+2q^2+3q^3+5q^4+7q^5+10q^6+...\right)
\end{eqnarray}

\end{widetext}

\section{Complexity}
\label{sec:noisy_ising}
\begin{figure}[t!]
\centering
\includegraphics[width=0.49\textwidth]{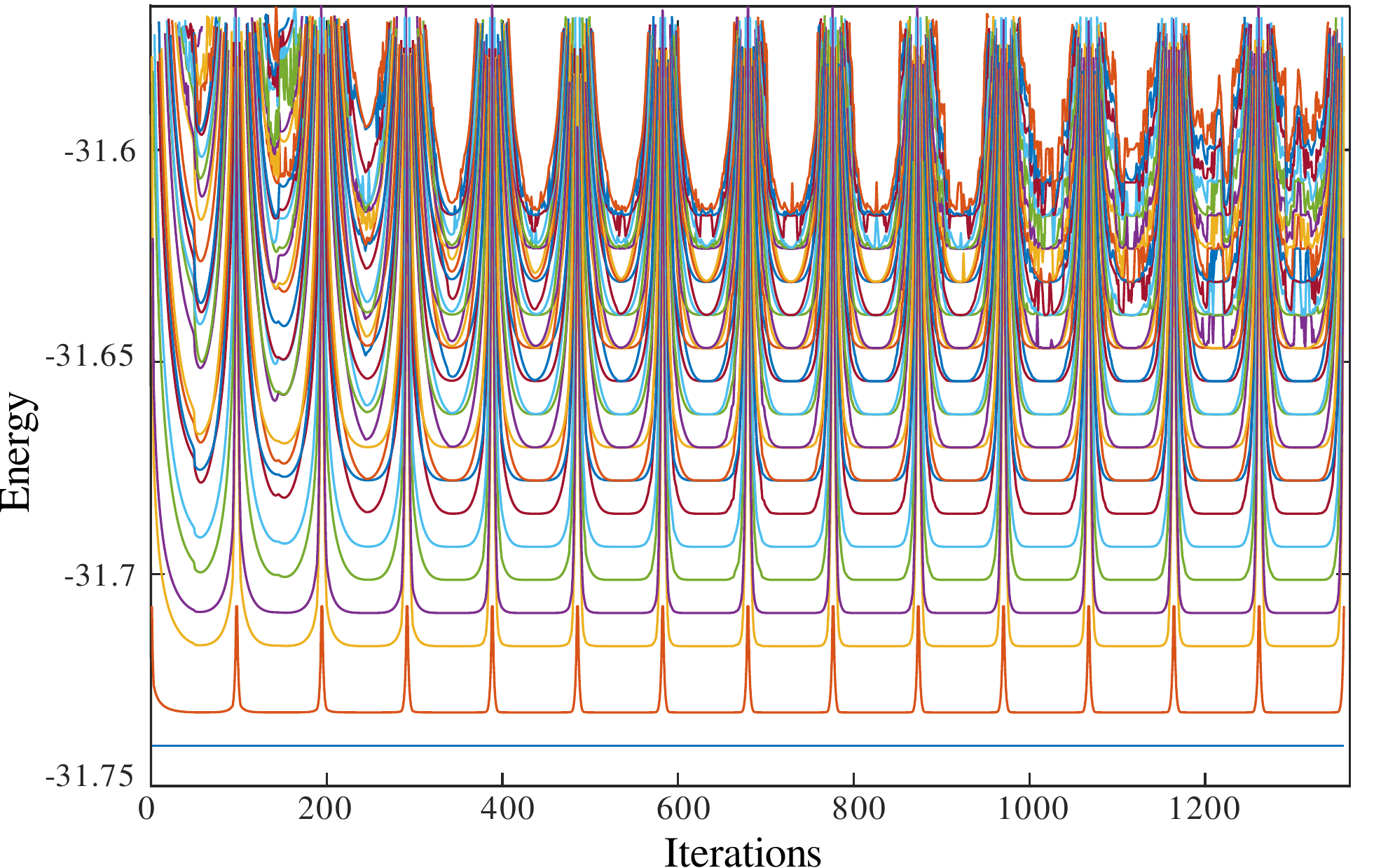}
\caption{(Color online) Same as Fig.\ref{fig:ising_conv} obtained with at most 200 Lanczos iterations. The number of states increases linearly from $D=50$ (in the warm-up) up to $D\approx 200$. When the number of Lanczos iterations is sufficiently large, the noise in the higher excited states is suppressed as shown in Fig.\ref{fig:ising_conv} obtained with at most 500 Lanczos iterations.}
\label{fig:noisy_ising}
\end{figure}

Some examples of convergence have already been shown for the critical Ising model in Fig.\ref{fig:ising_conv}  and for the three-state Potts model in Fig.\ref{fig:pot}.
We have noticed that the convergence of the excitation energies obtained by calculating many eigenvalues of the effective Hamiltonian depends not only on the number of kept states $D$, but is also extremely sensitive to the number of Lanczos iterations. In Fig.\ref{fig:noisy_ising} we show the convergence of the energy spectra in the critical Ising chain when the number of Lanczos iterations is restricted to 200, a typical number when looking for the ground state, as opposed to 500 in Fig.\ref{fig:ising_conv}. The noise that can 
be seen in the high-energy levels in Fig.\ref{fig:noisy_ising} disappears upon increasing the number of Lanczos iterations (Fig.\ref{fig:ising_conv}). The number of Lanczos iterations is the only parameter that increases the complexity of the algorithm when calculating the excitation spectrum as compared to the ground-state search.

\bibliographystyle{apsrev4-1}
\bibliography{bibliography}

\end{document}